\documentclass[aps,prb,twoside,superscriptaddress,longbibliography,nofootinbib,floatfix, notitlepage,twocolumn]{revtex4-1}

\usepackage[utf8]{inputenc}
\usepackage{color}
\usepackage{amsmath}
\usepackage{braket}
\usepackage{dsfont}
\usepackage[pdftex,colorlinks=true,linkcolor=blue,citecolor=blue,urlcolor=blue]{hyperref}
\usepackage{graphicx}
\usepackage{helvet}
\usepackage{sansmath} % to get figure captions with sans serif math

\usepackage{appendix}
\usepackage[caption=false,label font={bf},position=bottom]{subfig}
\usepackage{pgf}
\usepackage{tikz}
\usetikzlibrary{quantikz}
\DeclareGraphicsExtensions{.png,.pdf}

\usepackage[ruled,vlined]{algorithm2e}
\SetKwComment{Comment}{$\triangleright$\ }{}

\SetCommentSty{mycommfont}

\newcommand{\subcaption}[1]{\footnotesize \sffamily \sansmath #1}

\renewcommand{\vec}[1]{\boldsymbol{#1}}

\newcommand{\Nocc}{N_{\rm{occ}}}
\newcommand{\Noccsigma}{N_{\rm{occ},\sigma}}
\DeclareMathOperator{\tr}{tr}

\newcommand{\be}{\begin{equation}}
\newcommand{\ee}{\end{equation}}

\graphicspath{{pics/}}

\begin{document}

\title{Observing ground-state properties of the Fermi-Hubbard model using a scalable algorithm on a quantum computer}
\date{\today}
\author{Stasja Stanisic}
\affiliation{Phasecraft Ltd.}
\author{Jan Lukas Bosse}
\affiliation{Phasecraft Ltd.}
\affiliation{School of Mathematics, University of Bristol, UK}
\author{Filippo Maria Gambetta}
\affiliation{Phasecraft Ltd.}
\author{Raul A. Santos}
\affiliation{Phasecraft Ltd.}
\author{Wojciech Mruczkiewicz}
\affiliation{Google Quantum AI}
\author{Thomas E. O'Brien}
\affiliation{Google Quantum AI}
\author{Eric Ostby}
\affiliation{Google Quantum AI}
\author{Ashley Montanaro}
\affiliation{Phasecraft Ltd.}
\affiliation{School of Mathematics, University of Bristol, UK}

\begin{abstract}
The famous, yet unsolved, Fermi-Hubbard model for strongly-correlated electronic systems is a prominent target for quantum computers. However, accurately representing the Fermi-Hubbard ground state for large instances may be beyond the reach of near-term quantum hardware. Here we show experimentally that an efficient, low-depth variational quantum algorithm with few parameters can reproduce important qualitative features of medium-size instances of the Fermi-Hubbard model. We address $1\times 8$ and $2\times 4$ instances on 16 qubits on a superconducting quantum processor, substantially larger than previous work based on less scalable compression techniques, and going beyond the family of 1D Fermi-Hubbard instances, which are solvable classically. Consistent with predictions for the ground state, we observe the onset of the metal-insulator transition and Friedel oscillations in 1D, and antiferromagnetic order in both 1D and 2D. We use a variety of error-mitigation techniques, including symmetries of the Fermi-Hubbard model and a recently developed technique tailored to simulating fermionic systems. We also introduce a new variational optimisation algorithm based on iterative Bayesian updates of a local surrogate model. Our scalable approach is a first step to using near-term quantum computers to determine low-energy properties of strongly-correlated electronic systems that cannot be solved exactly by classical computers.
\end{abstract}

\maketitle

\setlength{\parskip}{3pt}

Understanding systems of many interacting electrons is a grand challenge of condensed-matter physics\cite{leblanc15}. This challenge is motivated both by practical considerations, such as the design and characterisation of novel materials\cite{scalapino2012}, and by fundamental science \cite{scalapino07,Arovas_review2021,Qin_review2021}. Yet classical methods are unable to represent the quantum correlations occurring in such systems efficiently, and accurately solving the many-electron problem for arbitrary large systems is beyond the capacity of the world's most powerful supercomputers.

This problem is thrown into sharp relief by the iconic Fermi-Hubbard model\cite{hubbard63,hubbard13}, the simplest system that includes non-trivial correlations not captured by classical methods (e.g.\ density functional theory). Although a highly simplified model of interacting electrons in a lattice, to date the largest Fermi-Hubbard system which has been solved exactly consisted of just 17 electrons on 22 sites\cite{yamada05}. Approximate methods can address much larger systems, but suffer from significant uncertainties in computing physically relevant quantities in certain regimes\cite{leblanc15}.

\tikzset{qubit/.style={shape=circle,fill=black, scale=0.5}}
\tikzset{qubit2/.style={shape=circle, draw=violet, scale=1.2}}
%\tikzset{->-/.style={decoration={markings, mark=at position #1 with {\arrow{latex}}},postaction={decorate}}}
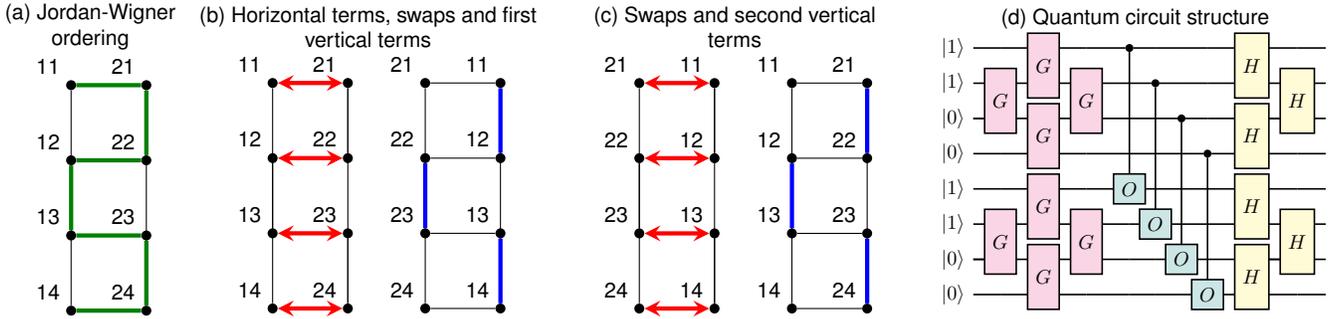
\begin{figure*}[t]
    \centering
    \begin{minipage}[c]{0.15\textwidth}
        \subcaption{(a) Jordan-Wigner ordering}\\
        \begin{tikzpicture}[node distance = 0.2cm]
            \foreach \x in {0,...,1}
            \foreach \y in {0,...,3}
            {
                \pgfmathtruncatemacro\xx{\x+1}
                \pgfmathtruncatemacro\yy{4-\y}
                \node[label={above left:\xx\yy},qubit] (\x-\y) at (\x,\y){};
            }
            \foreach \y in {0,...,3}
            {
                \draw[ultra thick, draw=green!50!black] (0-\y.east) -- (1-\y.west);
            }
            \draw (0-3.south) -- (0-0.north);
            \draw (1-3.south) -- (1-0.north);
            \draw[ultra thick, draw=green!50!black] (1-3.south) -- (1-2.north);
            \draw[ultra thick, draw=green!50!black] (1-1.south) -- (1-0.north);
            \draw[ultra thick, draw=green!50!black] (0-2.south) -- (0-1.north);
        \end{tikzpicture}
    \end{minipage}
    \begin{minipage}[c]{0.25\textwidth}
        \subcaption{(b) Horizontal terms, swaps and first vertical terms}\\
        \begin{tikzpicture}[node distance = 0.2cm]
            \foreach \x in {0,...,1}
            \foreach \y in {0,...,3}
            {
                \pgfmathtruncatemacro\xx{\x+1}
                \pgfmathtruncatemacro\yy{4-\y}
                \node[label={above left:\xx\yy},qubit] (\x-\y) at (\x,\y){};
            }
            \foreach \y in {0,...,3}
            {
                \draw[<->,>=stealth,ultra thick, draw=red] (0-\y.east) -- (1-\y.west);
            }
            \draw (0-3.south) -- (0-0.north);
            \draw (1-3.south) -- (1-0.north);
            \draw (1-3.south) -- (1-2.north);
            \draw (1-1.south) -- (1-0.north);
            \draw (0-2.south) -- (0-1.north);
        \end{tikzpicture}
        \hspace{0.25cm}
        \begin{tikzpicture}[node distance = 0.2cm]
            \foreach \x in {0,...,1}
            \foreach \y in {0,...,3}
            {
                \pgfmathtruncatemacro\xx{2-\x}
                \pgfmathtruncatemacro\yy{4-\y}
                \node[label={above left:\xx\yy},qubit] (\x-\y) at (\x,\y){};
            }
            \foreach \y in {0,...,3}
            {
                \draw (0-\y.east) -- (1-\y.west);
            }
            \draw (0-3.south) -- (0-0.north);
            \draw (1-3.south) -- (1-0.north);
            \draw[ultra thick, draw=blue] (1-3.south) -- (1-2.north);
            \draw[ultra thick, draw=blue] (1-1.south) -- (1-0.north);
            \draw[ultra thick, draw=blue] (0-2.south) -- (0-1.north);
        \end{tikzpicture}
    \end{minipage}
    \hfill
    \begin{minipage}[c]{0.25\textwidth}
        \subcaption{(c) Swaps and second vertical terms}\\
        \begin{tikzpicture}[node distance = 0.2cm]
            \foreach \x in {0,...,1}
            \foreach \y in {0,...,3}
            {
                \pgfmathtruncatemacro\xx{2-\x}
                \pgfmathtruncatemacro\yy{4-\y}
                \node[label={above left:\xx\yy},qubit] (\x-\y) at (\x,\y){};
            }
            \foreach \y in {0,...,3}
            {
                \draw[<->,>=stealth,ultra thick, draw=red] (0-\y.east) -- (1-\y.west);
            }
            \draw (0-3.south) -- (0-0.north);
            \draw (1-3.south) -- (1-0.north);
            \draw (1-3.south) -- (1-2.north);
            \draw (1-1.south) -- (1-0.north);
            \draw (0-2.south) -- (0-1.north);
        \end{tikzpicture}
        \hspace{0.25cm}
        \begin{tikzpicture}[node distance = 0.2cm]
            \foreach \x in {0,...,1}
            \foreach \y in {0,...,3}
            {
                \pgfmathtruncatemacro\xx{\x+1}
                \pgfmathtruncatemacro\yy{4-\y}
                \node[label={above left:\xx\yy},qubit] (\x-\y) at (\x,\y){};
            }
            \foreach \y in {0,...,3}
            {
                \draw (0-\y.east) -- (1-\y.west);
            }
            \draw (0-3.south) -- (0-0.north);
            \draw (1-3.south) -- (1-0.north);
            \draw[ultra thick, draw=blue] (1-3.south) -- (1-2.north);
            \draw[ultra thick, draw=blue] (1-1.south) -- (1-0.north);
            \draw[ultra thick, draw=blue] (0-2.south) -- (0-1.north);
        \end{tikzpicture}
    \end{minipage}
    \hfill
    \begin{minipage}[c]{0.3\textwidth}
        \subcaption{(d) Quantum circuit structure}
        \includegraphics[width=0.95\textwidth]{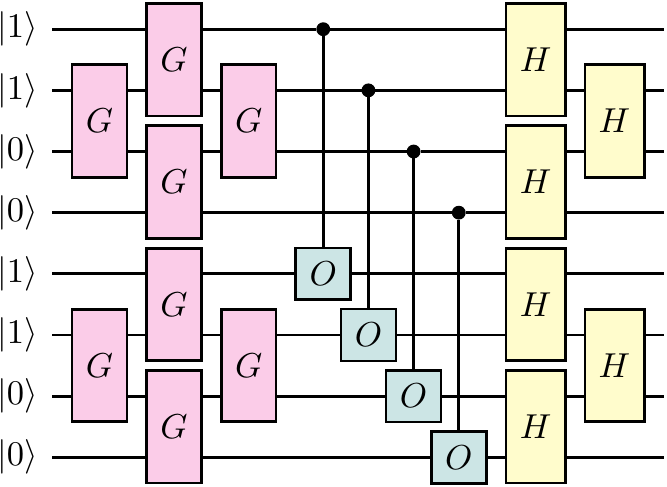}
    \end{minipage}
    \caption{Implementation of the Efficient Hamiltonian Variational ansatz. (a): Jordan-Wigner encoding mapping one spin sector of a $2\times 4$ lattice to a line. Mapping is repeated for the other spin sector. (b)--(c): Horizontal terms are implemented combined with fermionic swaps (red); then the first set of vertical terms (blue); then another layer of fermionic swaps; then the second set of vertical terms. (d): Quantum circuit structure shown for a $1 \times 4$ instance at half-filling with one variational layer (actual experiments used up to 16 qubits). $G$: Givens rotations; $O$: onsite gates; $H$: hopping gates. Onsite and hopping gates correspond to time-evolution according to onsite and hopping terms in the Fermi-Hubbard Hamiltonian; the structure of this part is repeated for multiple layers. All onsite terms have the same time parameter, and for $1 \times L_y$ instances, all hopping terms occurring in parallel have the same time parameter. When implemented on hardware in a zig-zag configuration, a layer of FSWAP gates is required before and after the onsite gates. First four qubits represent spin-up modes, last four represent spin-down modes. All operations in this diagram are implemented using two hardware-native 2-qubit gates. Circuit is repeated multiple times for energy measurement, with differing measurement transformations at the end.}
    \label{fig:2x4_swap}
\end{figure*}

Quantum computers can represent quantum systems natively, and may enable the solution of physical problems that classical computers cannot handle. The Fermi-Hubbard model has been widely proposed as an early target for quantum simulation algorithms\cite{wecker15strongly,wecker15,jiang2018quantum,cade20,cai20,dallairedemers20,martin2021variational}. As well as its direct application to understanding technologically-relevant correlated materials, the regularity and relative simplicity of the Fermi-Hubbard Hamiltonian suggest that it may be easier to solve using a quantum computer than, for example, a large unstructured molecule; on the other hand, the challenge that it presents for classical methods makes it an excellent benchmark for quantum algorithms.

Small-scale experiments have used quantum algorithms to find ground states of the Fermi-Hubbard model for instances on up to 4 sites\cite{linke18,montanaro20,suchsland21} using up to 4 qubits.
These experiments compress the model based on its symmetries; methods of this form, while having running time scaling polynomially with system size, are complex enough that solving a post-classical Fermi-Hubbard instance would not be viable on a near-term quantum computer.

Here we instead use an extremely efficient quantum algorithm, proposed in Ref.~[\onlinecite{cade20}] based on previous work\cite{wecker15,jiang2018quantum,kivlichan18}, to study medium-scale instances of the Fermi-Hubbard model without the need for compression. The algorithm fits within the framework of the variational quantum eigensolver\cite{peruzzo14,bharti2021noisy} (VQE) using the Hamiltonian variational ansatz\cite{wecker15}. Based on extensive classical numerics for Fermi-Hubbard instances on up to 12 sites\cite{cade20}, this algorithm may be able to find accurate representations of the ground state of Fermi-Hubbard instances beyond classical exact diagonalisation by optimising over quantum circuits where the number of ansatz layers scales like the number of sites, corresponding to several hundred layers of two-qubit gates. While substantially smaller than previous quantum circuit complexity estimates for post-classical simulation tasks, this is still beyond the capability of today's quantum computers.

In this work, we demonstrate that a far lower number of ansatz layers can nevertheless reproduce qualitative properties of the Fermi-Hubbard model on quantum hardware. We apply VQE to Fermi-Hubbard instances on $1\times 8$ and $2\times 4$ lattices, using a superconducting quantum processor\cite{sycamore}, and observe physical properties expected for the ground state, such as the metal-insulator transition, Friedel oscillations, decay of correlations, and antiferromagnetic order. These results rely on an array of error-mitigation techniques that improve substantially the accuracy of estimating observables on noisy quantum devices, opening the path to useful applications in the near future.

\section{Variational algorithm}

Our algorithms attempt to approximate the ground state of the Fermi-Hubbard model,
\be \label{eq:hubbard} H = -\sum_{\langle i, j \rangle,\sigma} (a_{i\sigma}^\dag a_{j\sigma} + a_{j\sigma}^\dag a_{i\sigma}) + U \sum_i n_{i\uparrow}n_{i\downarrow}, \ee
where $a_{i\sigma}$ $(a^\dagger_{i\sigma})$ is a fermionic operator that destroys (creates) a particle at site $i$ with spin $\sigma$, $n_{i\sigma}=a^\dagger_{i\sigma}a_{i\sigma}$ is the number (density) operator, and $\langle i, j \rangle$ denotes adjacent sites on a rectangular lattice.

Representing the Fermi-Hubbard Hamiltonian on a quantum computer requires a fermionic encoding. Here we use the well-known Jordan-Wigner transform, under which each fermionic mode maps to one qubit, interpreted as lying on a 1D line. This parsimony in space comes at the price that, except in 1D, some terms correspond to operators acting on more than two qubits:
\begin{align}
a_i^\dag a_j + a_j^\dag a_i &\mapsto \frac{1}{2}(X_iX_j + Y_iY_j) Z_{i+1} \cdots Z_{j-1},\\
%n_i n_j = a_i^\dag a_i a_j^\dag a_j &\mapsto \frac{1}{4}(I - Z_i)(I - Z_j).
n_i n_j = a_i^\dag a_i a_j^\dag a_j &\mapsto \proj{11}_{ij}.
\label{eq:qubitOps}
\end{align}
For $L_x \times L_y$ instances with $L_x \ge 2$, the ``snake'' ordering shown in Fig.\ \ref{fig:2x4_swap}(a) (for $2\times 4$) can be used to map the rectangular lattice to a line. Under this mapping, horizontal terms only involve pairs of qubits, but some vertical terms act on larger numbers of qubits. As onsite terms always only involve pairs of qubits, we can place the qubits corresponding to spin-down modes after those corresponding to spin-up without incurring any additional cost for these long-range interactions.

\begin{figure*}[t]
    \centering
    \includegraphics[width=.4\textwidth]{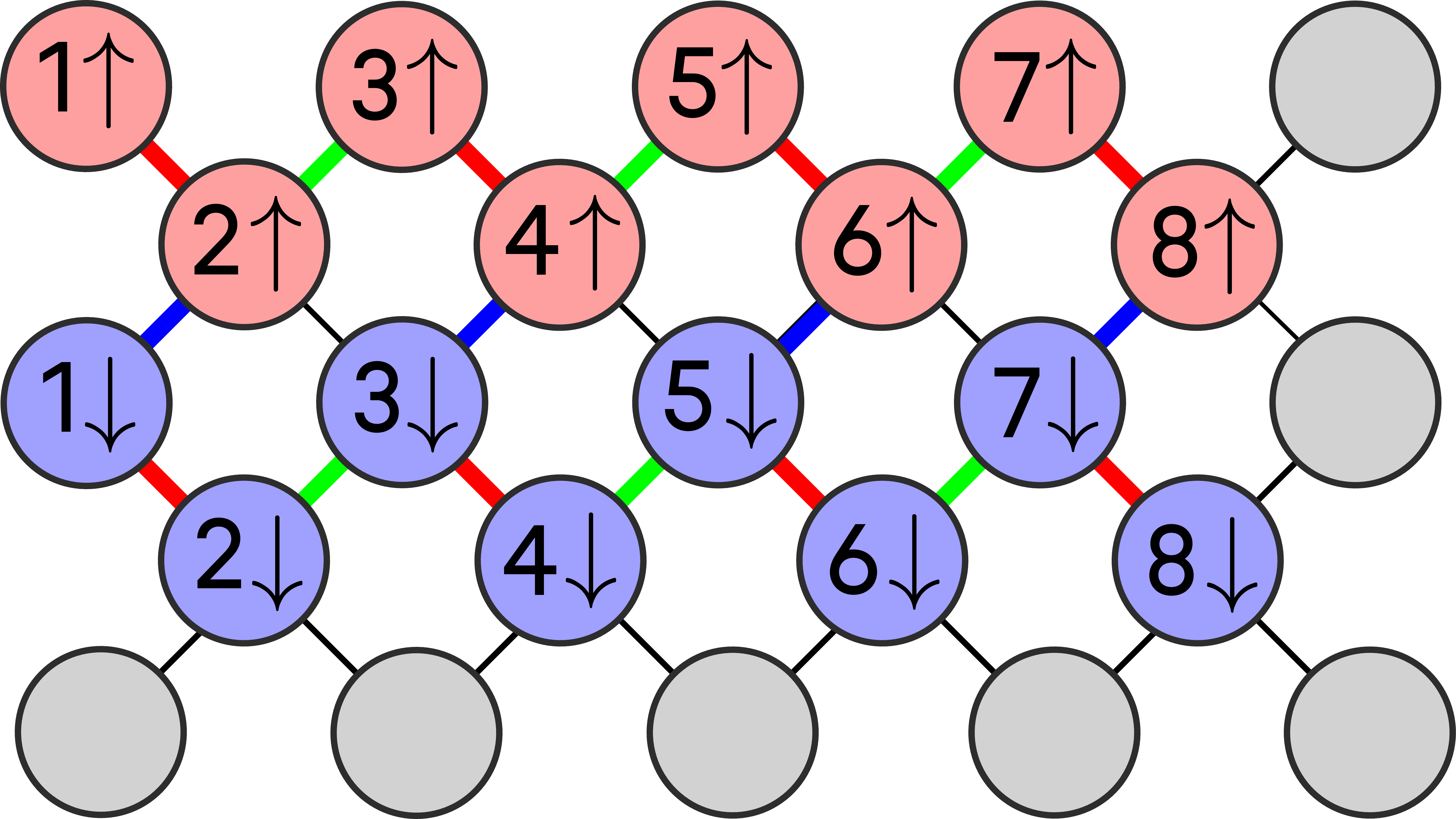}
    \hspace{2cm}
    \includegraphics[width=.4\textwidth]{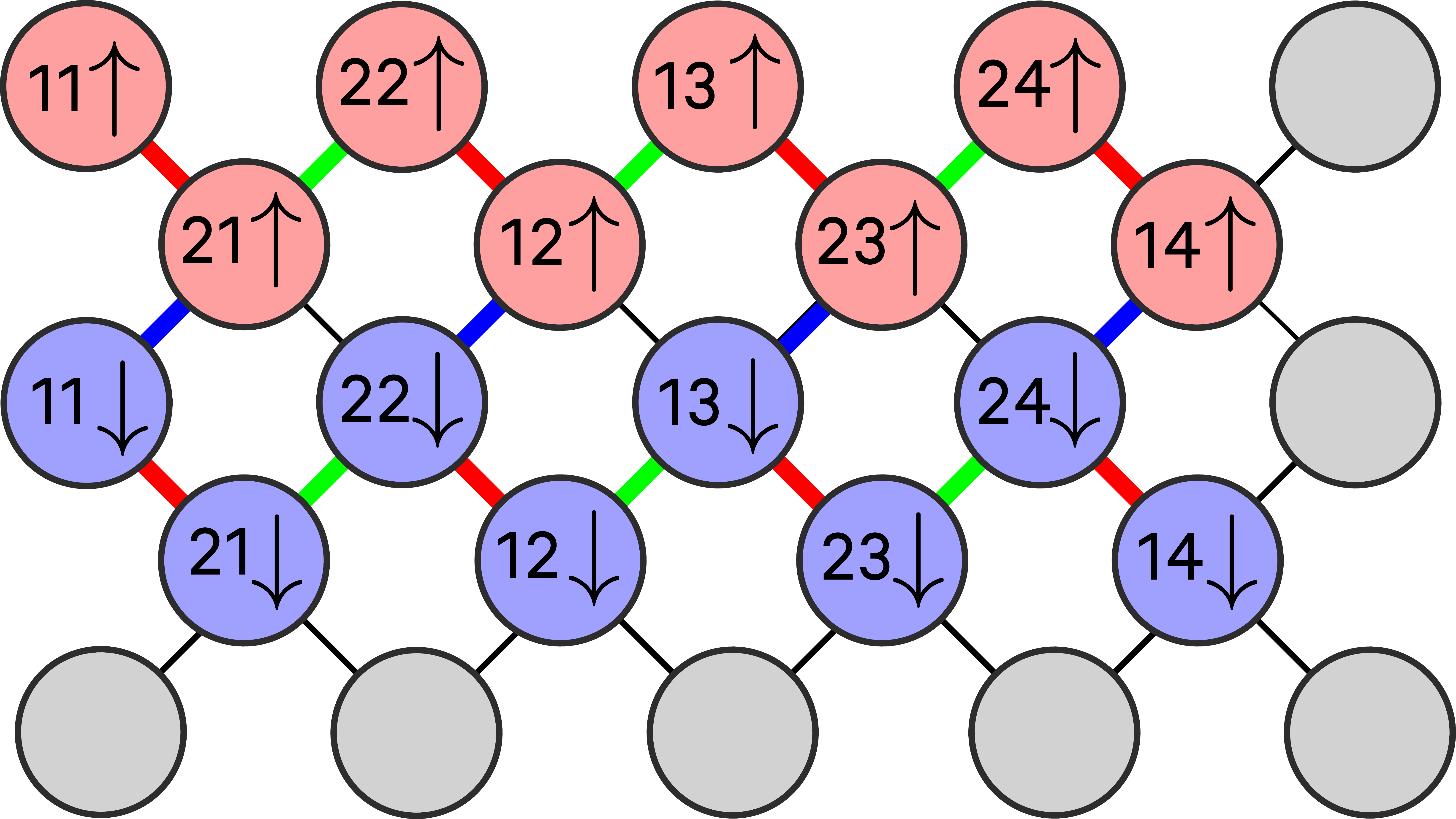}
    \caption{Qubit layout for implementing $1\times 8$ (left) and $2\times 4$ (right) Fermi-Hubbard instances. In each case two qubits are used to encode each site. Operations between qubits in variational layers occur in the following pattern. $1\times 8$: blue (FSWAP), red (onsite), blue (FSWAP), red (vert1), green (vert2). $2\times 4$: blue (FSWAP), red (onsite), blue (FSWAP), red (horiz+FSWAP), green (vert), red (FSWAP), green (vert). Vertical interaction parameters for $2\times 4$ depend on the parity of the row index. Grey circles denote the unused qubits on the 23-qubit Rainbow chip.}
    \label{fig:qubit_layout}
\end{figure*}

The variational approach we use optimises over quantum circuits of the following form\cite{cade20} (Fig.\ \ref{fig:2x4_swap}(d)). First, prepare the ground state of the noninteracting ($U=0$) Fermi-Hubbard model, which can be achieved efficiently via a sequence of Givens rotations\cite{jiang2018quantum}, which act on pairs of adjacent modes. Then repeat a number of layers, each consisting of time-evolution according to terms in the Fermi-Hubbard model.

The Hamiltonian $H$ has a natural decomposition into at most 5 sets of terms on a rectangular lattice
%(horizontal 1 and 2, vertical 1 and 2, and onsite)
such that all the terms in each set act on disjoint modes. This, in principle, allows the corresponding time-evolution steps to be implemented in parallel, although care must be taken over overlapping $Z$-strings in the Jordan-Wigner transform. Evolution times are variational parameters which are optimised using a classical optimisation algorithm. Within each layer, the terms within each set evolve for the same amount of time. For a $1 \times L_y$ instance, each layer then has 3 parameters (one onsite term, and two types of hopping terms); for a $2\times L_y$ instance, $L_y \ge 2$, each layer has 4 parameters; and for a $L_x\times L_y$ instance, $L_x,L_y \ge 3$, each layer has 5 parameters.

This structure is advantageous in two respects: the small number of parameters reduces the complexity of the variational optimisation process, and the variational ansatz respects the symmetries of the Fermi-Hubbard model, which (as we will see) provides opportunities for error mitigation. 
The same decomposition of $H$ into at most 5 parts allows for highly efficient measurement of energies using only 5 distinct measurements, each implemented via a computational basis measurement with at most one additional layer of 2-qubit gates\cite{cade20}.

The final component of the VQE framework is the classical optimisation routine that optimises over the parameters in the quantum circuit to attempt to minimise the energy, and hence produce the ground state. This optimisation process is challenging as measurements are noisy, due to statistical noise and to errors in the quantum hardware. Here we introduce a new algorithm for this optimisation procedure, which we call BayesMGD. It enhances the MGD (Model Gradient Descent) algorithm\cite{sung20,GoogleQAOA_2021} by performing iterative, Bayesian updates of a quadratic, local surrogate model to
the objective function to make optimal use of the information gained from noisy measurements at each time step of the algorithm. During each iteration, the prior knowledge of the local 
quadratic fit to the objective function is updated by evaluating the latter in a neighbourhood of the current parameters. The gradient of this improved quadratic fit is then used to perform a gradient descent step. See Appendix \ref{sec:experimental-comparison-of-optimisation-algorithms} for details of experimental results comparing BayesMGD, MGD and SPSA.

\section{Quantum circuit implementation}

We carried out our experiments on the ``Rainbow'' superconducting quantum processor in Google Quantum AI's Sycamore architecture, which had 23 qubits available in the configuration shown in Fig.\ \ref{fig:qubit_layout}.

We studied Fermi-Hubbard model instances on lattices of shapes $1\times L_y$ and $2\times L_y$. 
A $1 \times L_y$ Fermi-Hubbard system can be mapped to a $2 \times L_y$ rectangular lattice by associating each site with two adjacent qubits for spin-up and spin-down. All hopping and onsite interactions can be implemented locally, leading to a very efficient quantum circuit. However, on the hardware platform we used, this configuration would only support a lattice of size at most $1 \times 4$. To enable us to study systems of size up to $1 \times 8$, we used a ``zig-zag'' configuration consisting of two nearby lines of length 8 (Fig.\ \ref{fig:qubit_layout}). Hopping interactions are implemented as local operations within each line, but onsite interactions are non-local and require a layer of swap operations.

For a $2 \times L_y$ lattice, due to the Z-strings occurring in the Jordan-Wigner transform, implementing some of the vertical hopping interactions directly would require 4-qubit operations. To remove the need for these, we use a fermionic swap network\cite{kivlichan18}. A fermionic swap (FSWAP) operation rearranges the Jordan-Wigner ordering such that operations that were previously long-distance can now be implemented via 2-qubit gates. Here, swapping across the horizontal direction of the lattice allows vertical interactions to be implemented efficiently (Fig.\ \ref{fig:2x4_swap}). The overhead for a $2\times L_y$ lattice is only one additional layer of fermionic swap gates per layer of the variational ansatz, together with some additional fermionic swaps for measurement. However, using the fermionic swap network approach does restrict the order in which terms are implemented, as vertical interactions occur across pairs determined by the Jordan-Wigner ordering. We therefore give this variational ansatz a specific name, the Efficient Hamiltonian Variational ansatz\cite{cade20}.

In terms of quantum circuit complexity, the most complex instances we address are at or near half-filling, where with one variational layer, a $1\times 8$ instance requires 2-qubit gate depth at most 26 and at most 140 2-qubit gates, and a $2 \times 4$ instance requires 2-qubit gate depth at most 32 and at most 176 2-qubit gates. For further implementation details, see Appendix \ref{app:ansatz}.

\begin{figure*}
    \centering
    \begin{minipage}[c]{0.48\linewidth}
        \subcaption{(a) Progress of VQE for $1\times 8$ and $2\times 4$ Fermi-Hubbard instances, $U=4$, at half-filling}
        \includegraphics[scale=0.95]{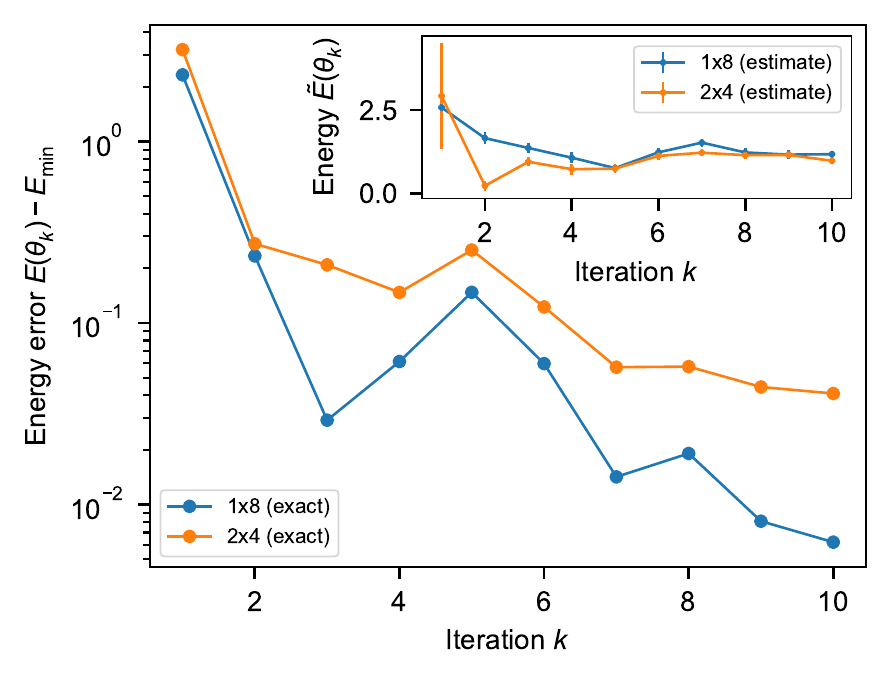}
    \end{minipage}
    \begin{minipage}[c]{0.48\linewidth}
        \subcaption{(b) Energy errors for $1\times 8$, $U=4$}
        \includegraphics[scale=0.95]{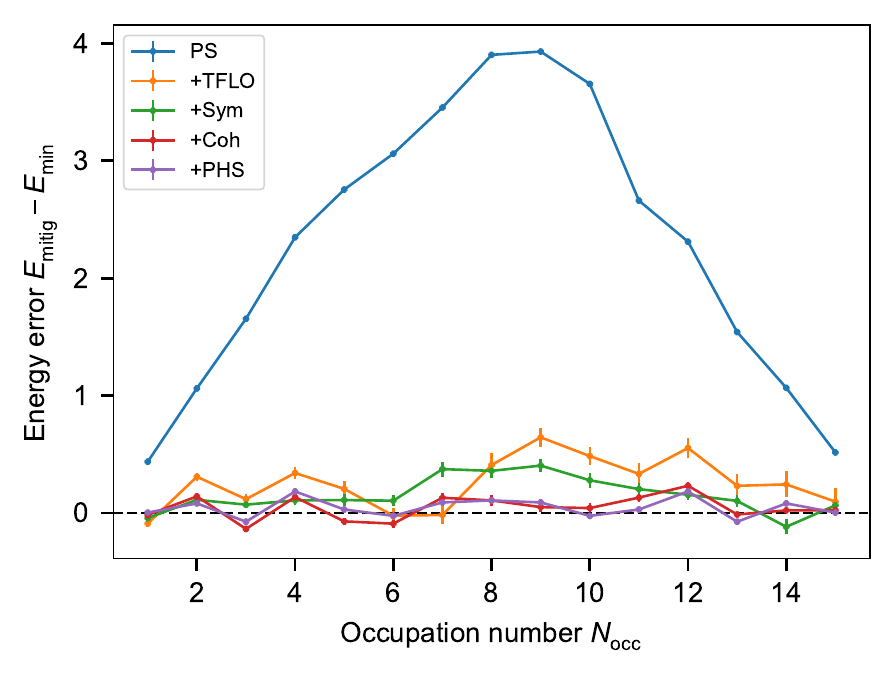}
    \end{minipage}
    \caption{Experimental results for the BayesMGD algorithm and final energy errors. (a) Progress of VQE for $1\times 8$ and $2\times 4$ Fermi-Hubbard instances at half-filling, as measured by the error between energy at parameters $\mathbf{\theta}_k$ and VQE ground energy $E_{\min}$ (main plot log scale, inset linear scale). ``Estimate'' is the energy estimated by the BayesMGD algorithm during the VQE procedure based on measurement results, ``exact'' is the true energy at the corresponding parameters. (b) Final errors in measured energy following error mitigation on the final state. ``PS'': only postselection on occupation number. ``Sym'': also time-reversal symmetry. ``TFLO'': also Training with Fermionic Linear Optics\cite{montanaro21}. ``Coh'': also coherent error correction in TFLO. ``PHS'': also particle-hole symmetry. Each error mitigation method is applied as well as all previous methods. Plots show a piecewise linear interpolation between integer occupations. Error bars were calculated according to the procedure in Appendix \ref{app:error_analysis} and are often too small to be visible.}
    \label{fig:vqe_progress}
\end{figure*}

\section{Error mitigation}

Achieving accurate results requires a variety of error-mitigation procedures, divided into three categories. First, we use low-level circuit optimisations tailored to the hardware platform. Second, we take advantage of the symmetries of the Fermi-Hubbard Hamiltonian. And finally we use a technique for mitigating errors in fermionic Hamiltonian simulation algorithms.

We begin by optimising the quantum circuit to contain alternating layers of 1-qubit and 2-qubit gates, and selecting a high-performance set of qubits to use based on an initial test. We then use a technique based on spin-echo\cite{freeman98} where every other layer of 2-qubit gates is sandwiched between layers of X gates on every qubit. This led to a substantial reduction in error in our experiments, which we attribute to two possible causes: that these X gates are inverting single-qubit phase errors that accumulate during the circuit; and that they modify ``parasitic CPHASE'' errors occurring on the 2-qubit gates, which are known to be substantial\cite{GoogleFH_TDS_2020}.

The symmetry-based techniques for error mitigation that we use exploit number conservation per spin sector, time reversal, particle-hole and lattice symmetries. Number conservation allows error-detection by discarding runs where final and initial occupations do not match.  In particular, this detects many errors that occur due to incorrect qubit readout, a significant source of error in superconducting qubit systems. The other three symmetries allow us to average results obtained from a state and its symmetry-transformed partner.

The last error-mitigation technique we used is targeted at quantum algorithms for general fermionic systems\cite{montanaro21}, and called Training with Fermionic Linear Optics (TFLO). TFLO uses efficient classical simulation of quantum circuits of time-evolution operations by quadratic fermionic Hamiltonians\cite{terhal02} (so-called fermionic linear optics (FLO) circuits). Expectations of energies, or other observables of interest, for states produced by FLO circuits can be calculated exactly classically, and approximately using the quantum computer. These pairs of exact and approximate energies can be used as training data to infer a map from approximate energies computed by the quantum computer, at points which are not accessible classically, to exact energies. For this map to be accurate, the FLO circuits should approximate the real circuits occurring in the algorithm.

\begin{figure*}
    %\centering
%    \begin{minipage}{\textwidth}
    \begin{minipage}[c]{0.32\linewidth}
    \subcaption{(a) Energies, $1\times8$, $U=4$}
    \includegraphics[scale=0.95]{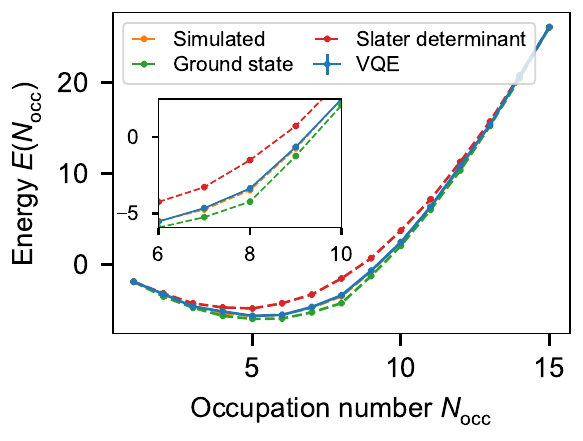}
    \end{minipage}
    \begin{minipage}[c]{0.32\linewidth}
    \subcaption{(b) Energies, $2\times4$, $U=4$}
    \includegraphics[scale=0.95]{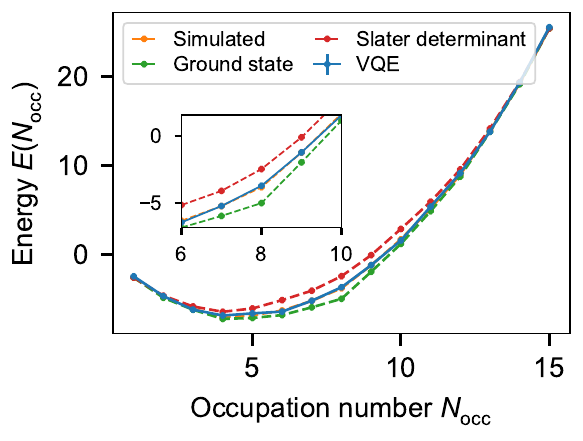}
    \end{minipage}
    \begin{minipage}[c]{0.32\linewidth}
    \subcaption{(c) Energies, $1\times4$, $U=4$}
    \includegraphics[scale=0.95]{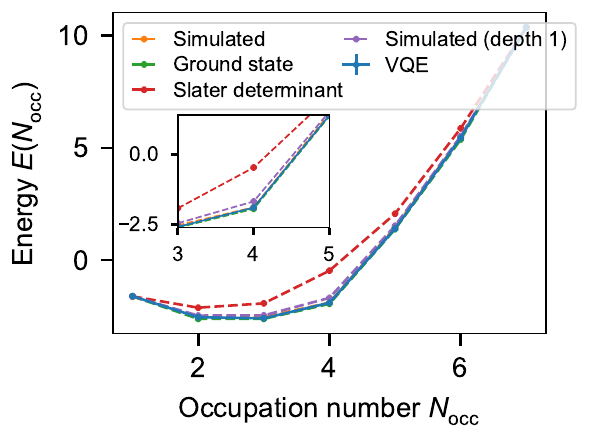}
    \end{minipage}\\
    \begin{minipage}[c]{0.32\linewidth}
    \subcaption{(d) Chemical potentials, $1\times8$, $U=4$}
    \includegraphics[scale=0.95]{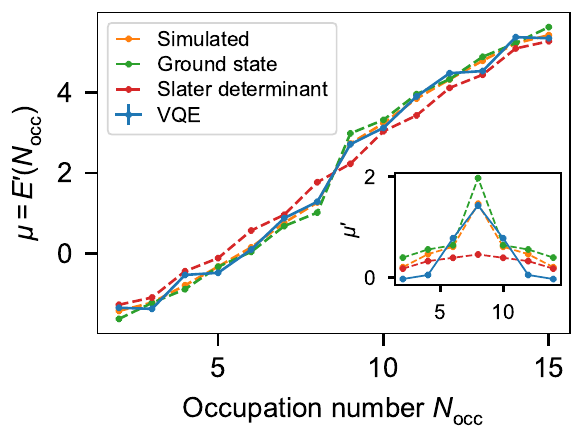}
    \end{minipage}
    \begin{minipage}[c]{0.32\linewidth}
    \subcaption{(e) Chemical potentials, $1\times8$, $U=8$}
    \includegraphics[scale=0.95]{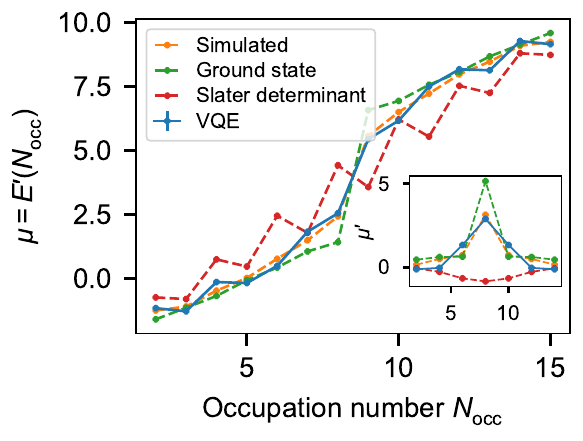}
    \end{minipage}
    \begin{minipage}[c]{0.32\linewidth}
    \subcaption{(f) Charge correlations, $1\times 8$, $U=4$}
    \includegraphics[scale=0.95]{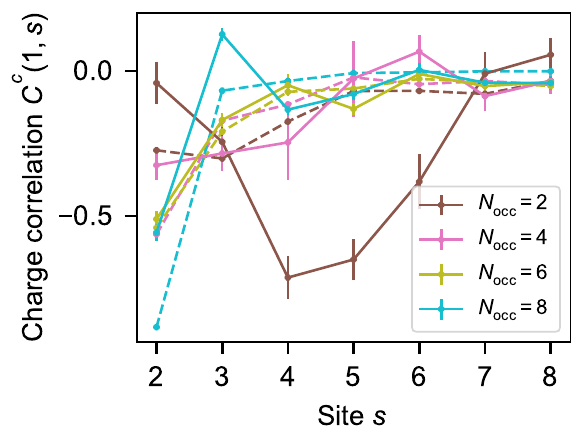}
    \end{minipage}
    \caption{Experimental energies, chemical potentials and charge correlations. ``VQE'': experimental data. ``Simulated'': the lowest energy achievable in the VQE ansatz. ``Ground state'': energy in the true ground state within each occupation number subspace. ``Slater determinant'': the energy achieved by an optimised Slater determinant state as detailed in Appendix \ref{app:data_analysis}. Dashed lines are exact numerical calculations, solid line is experimental data. Plots show a piecewise linear interpolation between integer occupations. (a)--(c): Energies $E(\Nocc)$ produced by VQE experiments compared with exact results ($U=4$). VQE results for $1\times8$ and $2\times4$ use one variational layer; $1\times4$ has two variational layers. Inset shows zoomed-in region around half-filling. (d)--(e): Chemical potentials $\mu$ for a $1\times8$ system, where $\mu(\Nocc) = E(\Nocc)-E(\Nocc-1)$. Inset shows the derivative $\mu'(\Nocc) = E(\Nocc+1)-2E(\Nocc)+E(\Nocc-1)$ of the chemical potential at even occupations. (f): Decay of normalised charge correlations $C^c(1,i) = (\braket{n_1 n_i} - \braket{n_1}\braket{n_i}) / (\braket{n_1^2} - \braket{n_1}^2)$ for even occupation numbers. Solid lines: experimental results. Dashed lines: correlations in ground state.  Error bars were calculated according to the procedure in Appendix \ref{app:error_analysis} and are often too small to be visible.}
    \label{fig:energies}
\end{figure*}

TFLO is ideally suited to the Fermi-Hubbard model, as most of the operations in the VQE circuit are FLO operations, including initial state preparation, time-evolution by the hopping terms, and measurement transformations. The only operations in the circuit that are not FLO are time-evolution by the onsite terms. Therefore, we can find a suitable training set by choosing arbitrary parameters for the hopping terms and setting the parameters of the onsite terms to 0. 
Compared with previous implementations\cite{montanaro21}, here we improve accuracy by choosing these parameters carefully to maximise their spread, using a linear fitting algorithm designed to handle outliers\cite{theil50,sen68}, and implementing a final step to subtract off residual error.
More details on all our error mitigation techniques are included in Appendix \ref{app:error_mitigation}, and results are shown in Fig.~\ref{fig:vqe_progress}(b).

\begin{figure*}
    %\centering
    \hspace{-2cm}
    \begin{minipage}[c]{0.35\linewidth}
    \subcaption{(a) VQE, $U=4$}
    \includegraphics[height=5cm]{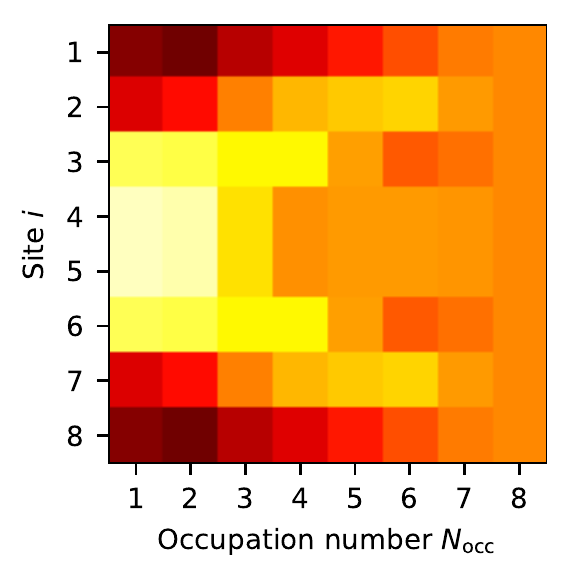}\\
    \subcaption{(e) Ground state, $U=4$}
    \includegraphics[height=5cm]{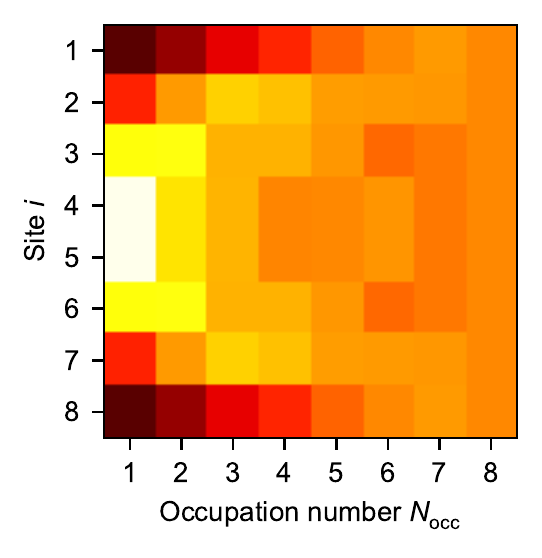}
    \end{minipage}
    \hspace{-0.8cm}
    \raisebox{0.5cm}{\begin{minipage}[l]{0.02\linewidth}
    \includegraphics[height=9.5cm]{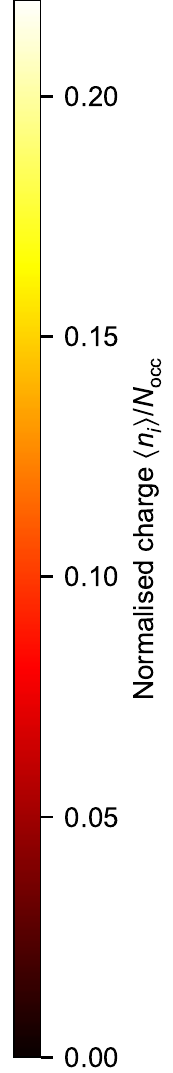}
    \end{minipage}}
    \hspace{2cm}
    \begin{minipage}[l]{0.15\linewidth}
    \subcaption{(b) VQE, $U=4$, $\Nocc$ even}
    \includegraphics[height=5cm]{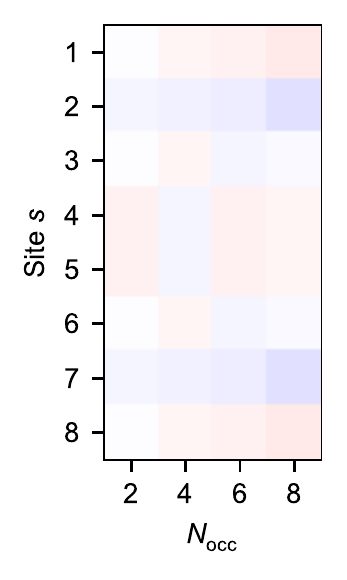}
    \subcaption{(f) Ground state, $U=0$, $\Nocc$ odd}
    \includegraphics[height=5cm]{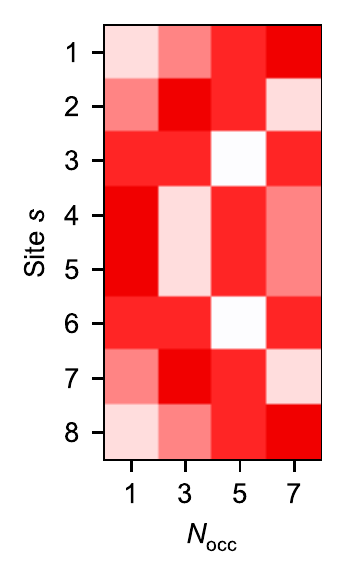}
    \end{minipage}
    \hspace{0.2cm}
    \begin{minipage}[l]{0.15\linewidth}
    \subcaption{(c) VQE, $U=4$, $\Nocc$ odd}
    \includegraphics[height=5cm]{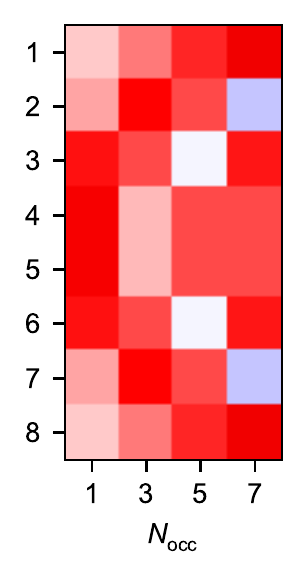}\\
    \subcaption{(g) Ground state, $U=4$, $\Nocc$ odd}
    \includegraphics[height=5cm]{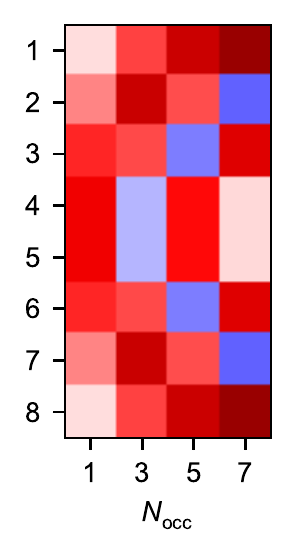}
    \end{minipage}
    \begin{minipage}[l]{0.15\linewidth}
    \subcaption{(d) VQE, $U=8$, $\Nocc$ odd}
    \includegraphics[height=5cm]{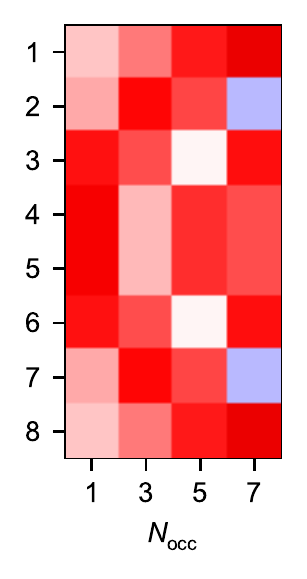}\\
    \subcaption{(h) Ground state, $U=8$, $\Nocc$ odd}
    \includegraphics[height=5cm]{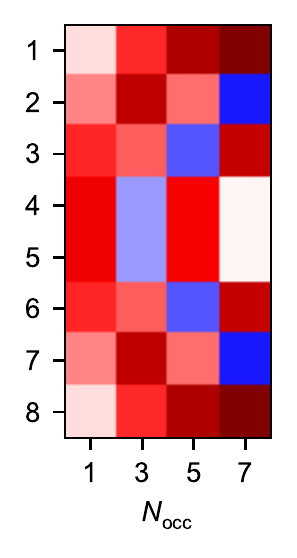}
    \end{minipage}
    \raisebox{0.8cm}{\begin{minipage}[l]{0.02\linewidth}
    \includegraphics[height=9.5cm]{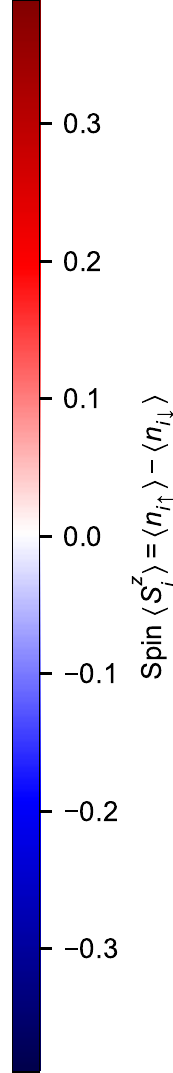}
    \end{minipage}}
    \caption{Charge and spin densities for a $1\times 8$ lattice. Top row: Experimental VQE results. Bottom row: Exact ground state. X axis: occupation number $\Nocc$; Y axis: site index. Spin plots split by even/odd occupations. In the ground state, spin is 0 everywhere for $\Nocc$ even. 
    }
    \label{fig:densities_spins}
\end{figure*}

\begin{figure*}
    %\centering
    \begin{minipage}[c]{0.49\linewidth}
    \subcaption{(a) $1\times 8$, $U=4$}
    \includegraphics[width=0.8\linewidth]{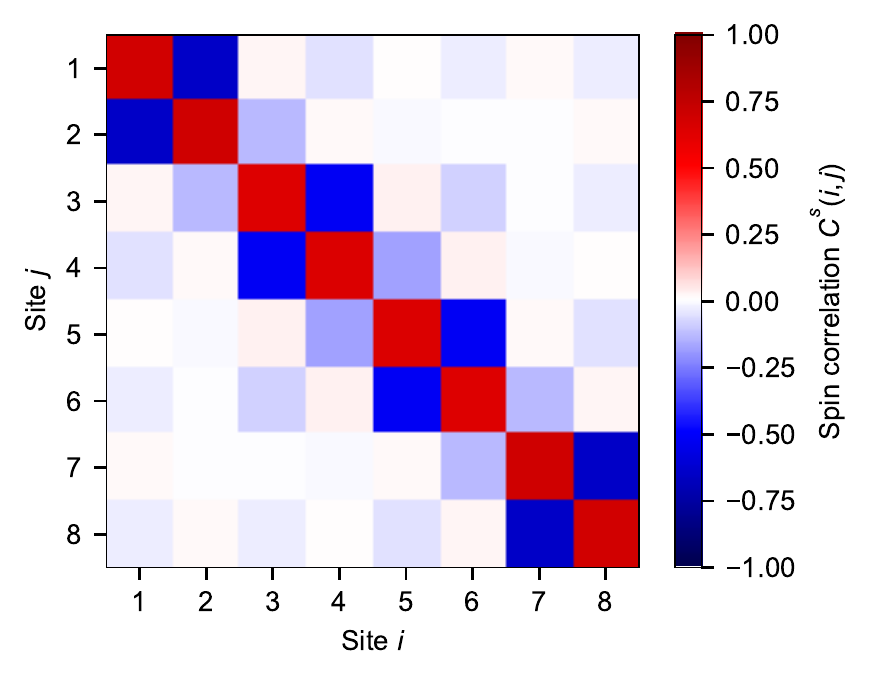}
    \end{minipage}
    \begin{minipage}[c]{0.49\linewidth}
    \subcaption{(b) $1\times 8$, $U=4$}
    \includegraphics[width=0.8\linewidth]{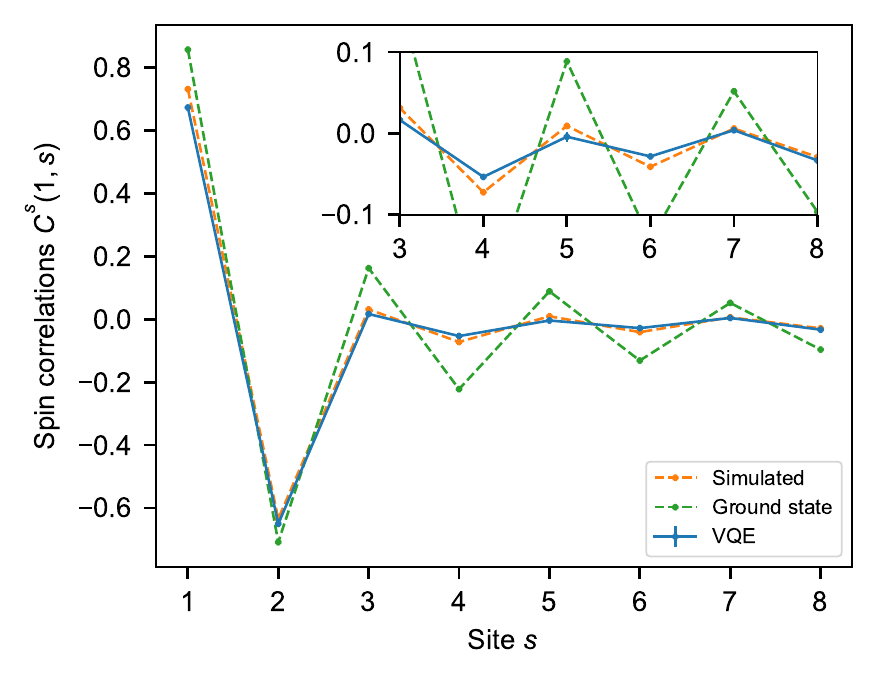}
    \end{minipage}
    \begin{minipage}[c]{0.49\linewidth}
    \subcaption{(c) $2\times 4$, $U=4$}
    \includegraphics[width=0.8\linewidth]{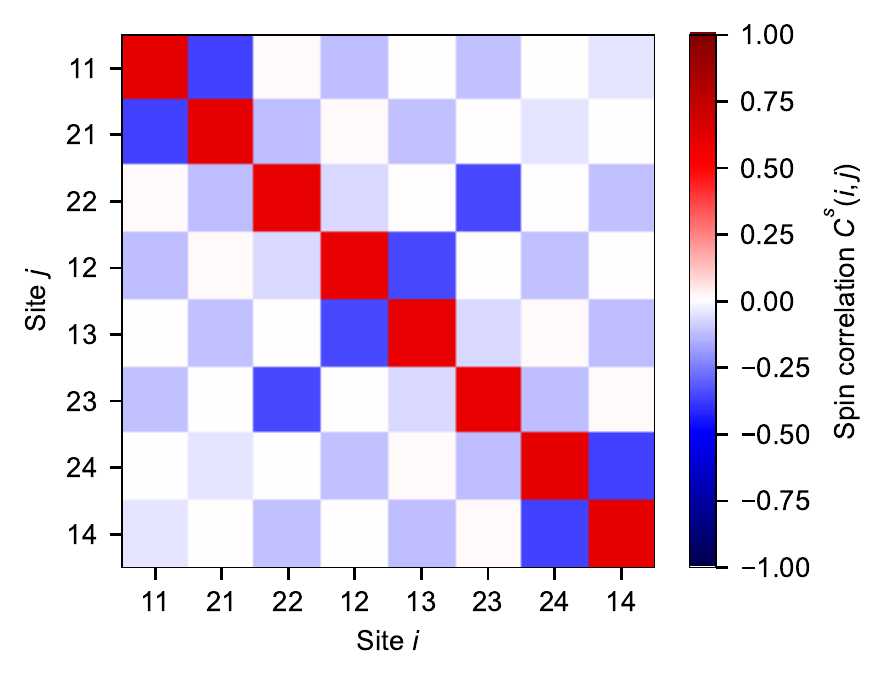}
    \end{minipage}
    \begin{minipage}[c]{0.49\linewidth}
    \subcaption{(d) $2\times 4$, $U=4$}
    \includegraphics[width=0.8\linewidth]{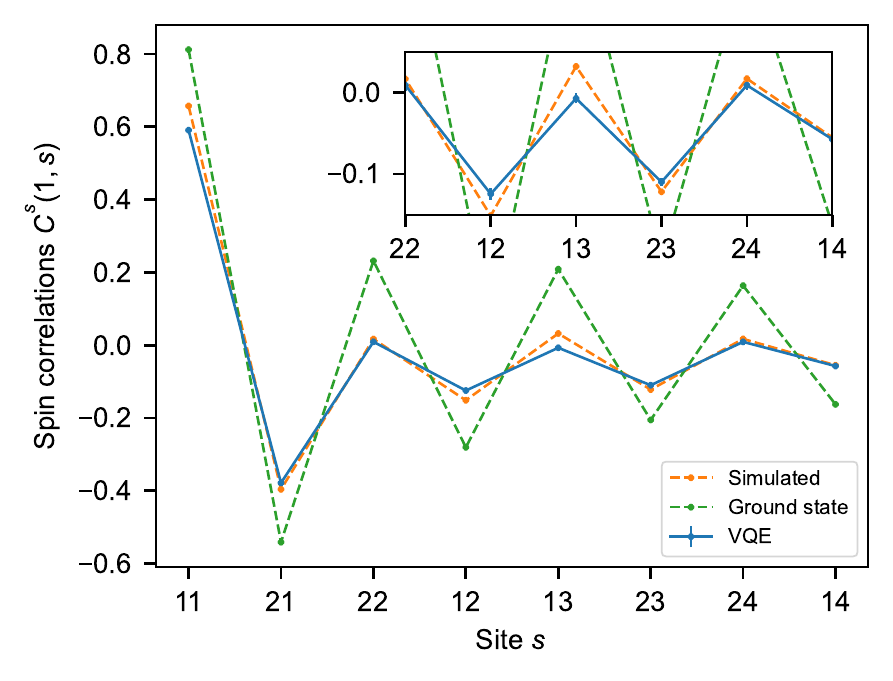}
   \end{minipage}
    \caption{Antiferromagnetic correlations at half-filling ($U=4$) obtained with the quantum processor. Ordering of sites for the $2\times 4$ lattice follows the Jordan-Wigner ``snake'' (Fig.\ \ref{fig:2x4_swap}). Meaning of labels in panels (b), (d) is as in Fig.\ \ref{fig:energies}.
    }
    \label{fig:spin_correlations}
\end{figure*}

\section{Results}

We used the BayesMGD algorithm within the VQE framework to determine the parameters required to produce approximate ground states of instances of the Fermi-Hubbard model on up to 8 sites, by minimising the energy expectation value calculated from the state produced by the VQE circuit on the quantum processor.
Once these parameters are determined, we have a quantum circuit to produce this state -- which we call the \emph{VQE ground state} below -- and can perform measurements to determine its properties. We found that BayesMGD was able to converge on parameters corresponding to the VQE ground state within a small number of iterations (Fig.\ \ref{fig:vqe_progress}).

First we compute the energy in the VQE ground state for $1\times 8$, $2\times 4$ and $1 \times 4$ systems for all occupation numbers $1 \le \Nocc \le 15$ ($\le7$ for the $1 \times 4$ system) (Fig.\ \ref{fig:energies}). In all cases good quantitative agreement is achieved with the exact lowest energy achievable with 1 layer of the VQE ansatz. To validate that the quantum algorithm goes beyond what is achievable with a straightforward classical ansatz, we compare with energies achieved by optimised Slater determinant states (Appendix \ref{app:data_analysis}). Further, in the $1\times 4$ case (Fig.\ \ref{fig:energies}(c)), lower energy is achieved with a depth 2 variational ansatz than is theoretically possible with depth 1, demonstrating that increased ansatz depth can lead to higher performance.

Next, we study the onset of the metal-insulator transition (MIT) \cite{Imada1998} between half-filling and away from half-filling in a $1\times 8$ system (Fig.~\ref{fig:energies}). Although in finite systems there is no true phase transition, we concentrate on two signals that are a precursor to this transition. First, a Mott gap which increases with $U$, shown by a nonzero derivative of the chemical potential (i.e., the second derivative of the energy, here approximated as $E(\Nocc+1)+E(\Nocc-1)-2E(\Nocc)$) at half-filling ($\Nocc=8$), when $U\neq0$ (see insets in Figs.~\ref{fig:energies} (d)-(e)). The physical origin of this can be understood as the energy penalty imposed for adding an electron on top of a half-filled state, where all sites are occupied. While in a 1D system of size $L_x$ the energy difference between states with occupations away from half-filling scales as $O(1/L_x)$, a fixed gap to charged excitations is a unique characteristic of a Mott insulator.
Second, we observe the spatial decay of normalised charge correlations with distance from the first site, $C^c(1,i) := (\braket{n_1 n_i} - \braket{n_1}\braket{n_i}) / (\braket{n_1^2} - \braket{n_1}^2)$ (Fig. \ref{fig:energies} (f)), where $n_i=n_{i\uparrow}+n_{i\downarrow}$. The steepest decay appears at half-filling ($\Nocc=8$), where the Mott gap implies the exponential decay of correlations. Further away from half-filling, the slower decay is a signature of increased conductivity. We have also computed these quantities for a $2\times 4$ system, where the results are suggestive but the MIT is not clear (see Appendix \ref{app:2x4_figures}).

Following, we study the behaviour of charge and spin densities at different sites and occupation numbers (Fig.~\ref{fig:densities_spins}). Boundaries in a finite-size system break the translational invariance and, as a consequence, induce Friedel oscillations in the charge density of the ground state\cite{Friedel1958} with twice the Fermi wavevector $k_F$. Therefore, in a 1D system with even (odd) occupation number $\Nocc$, they result in a ground-state charge density profile with $\Nocc/2$ ($(\Nocc+1)/2$) peaks. Evidence of this behaviour can be clearly seen in the VQE results in Fig.\ \ref{fig:densities_spins}(a). On the other hand, for strong onsite interactions and/or low fillings Wigner oscillations with wavevector $ 4k_F $ are also expected as a consequence of the Coulomb repulsion\cite{Schulz1993, Soffing2009}. In 1D, the latter are responsible for the emergence of $\Nocc$ peaks in the ground-state charge density and are indeed visible in Fig.\ \ref{fig:densities_spins}(e), especially for $ \Nocc \leq 4$. Hence, a comparison between Figs. \ref{fig:densities_spins}(a) and \ref{fig:densities_spins}(e) suggests that a higher-depth variational ansatz is required to fully capture strong interaction effects.
We see that, following error mitigation, the density in the case of equal number of spin-up and spin-down electrons is indeed close to zero (Fig.\ \ref{fig:densities_spins}(b)) as expected from symmetry, compared with the more substantial densities for odd occupations (Fig.\ \ref{fig:densities_spins}(c)--(d)), which in our case always corresponds to including an extra spin-up particle. These densities display a similar structure to the charge densities at the corresponding occupation.

To explore the differences between 1D and 2D, we compute (Fig.\ \ref{fig:spin_correlations})  the spin correlations $C^s(i,j) := \braket{S^z_i S^z_j} - \braket{S^z_i}\braket{S^z_j}$, where $S^z_i = n_{i\uparrow} - n_{i\downarrow}$, in the VQE ground state at half-filling for $1\times 8$ and $2\times 4$ lattices with strong onsite interaction ($U=4$).
We observe antiferromagnetic correlations compatible with the expected behaviour for that size, which are stronger compared with 1D. Antiferromagnetic and charge-density-wave ordering around half-filling are expected features of the Mott state in 2D. The charge profile for the $2\times 4$ system is reported in Fig.\ \ref{fig:2x4_densities_spins} of the Appendix (see also discussion therein). We also explore the antiferromagnetic character of the ground state for different onsite interactions and for occupations 7 and 8 (Fig.\ \ref{fig:mott2x4}) %in the light of the TNT result, 
where we observe that the system is indeed less antiferromagnetic at $\Nocc =7$ than at half-filling for $U=4,8$. Although the VQE ground state does not capture the value of the total staggered spin correlation in the true ground state quantitatively, it does follow the same trend.

\section{Discussion}

We have shown that fundamental qualitative features of medium-size instances of the Fermi-Hubbard model, using a number of qubits 4 times larger than Fermi-Hubbard experiments previously reported in the literature, can be extracted using a quantum computer with a low-depth variational ansatz. To do this, we achieve a relatively high level of accuracy in computing energies for states that can be produced with one variational layer. We expect that the use of a higher-depth variational ansatz in larger systems will enable the demonstration of phenomena such as Wigner oscillations, charge-density-wave ordering, and magnetic instabilities, and will shed some light on the different phases of the 2D system. Achieving a high level of quantitative accuracy in computing true ground state energies is a more significant challenge, which we expect will require a larger number of variational layers still, scaling with the system size\cite{cade20}. Our efficient algorithm and error-mitigation techniques provide a template that can readily be scaled up to larger systems as quantum computing hardware continues to improve.

\section*{Acknowledgements}

This project has received funding from the European Research Council (ERC) under the European Union's Horizon 2020 research and innovation programme (grant agreement No.\ 817581) and from EPSRC grant EP/S516090/1. Google Cloud credits were provided by Google via the EPSRC Prosperity Partnership in Quantum Software for Modeling and Simulation (EP/S005021/1). We would like to thank members of the Phasecraft and Google Quantum AI teams for helpful suggestions, and in particular Toby Cubitt, Ryan Babbush, Yu Chen, Charles Neill and Pedram Roushan. We would also like to thank Benjamin Chiaro, Brooks Foxen, and Kevin Satzinger for their work on building and maintaining the Rainbow processor.

\bibliographystyle{mybibstyle}
\bibliography{main} %,strategies}

\newpage
\appendix

\section*{Supplementary material}

%-------------------------------------------

\section{Physics of the Fermi-Hubbard model}

The Fermi-Hubbard model \cite{hubbard63} encapsulates the effect of inter-electron interactions in a single narrow band system. 
It was proposed to describe $d$-orbitals in transition metals.
In these systems, the strength of the interactions between localised electron orbitals is comparable with the bandwidth. This makes it a natural candidate for studying strongly correlated effects in solids \cite{DMFTreview}. 
The phenomenology of different cuprates that develop high-temperature superconductivity (e.g. $\rm{Sr}_{1-x}\rm{Ca}_x\rm{CuO}_2$) has been linked with the different phases of the model, although is still open if the model itself can support a high temperature superconducting phase \cite{Arovas_review2021}.

A general tight-binding description of the electron Hamiltonian in solids is
\begin{equation}
    H=-\sum_{\alpha\beta}t_{\alpha\beta}a^\dagger_\alpha a_\beta+\sum_{\alpha\beta\gamma\delta}U_{\alpha\beta\gamma\delta}a^\dagger_\alpha a^\dagger_\beta a_\gamma a_\delta,
\end{equation}
where $t_{\alpha\beta}$ contains the contributions from the kinetic energy and ion potential of the lattice, while $U_{\alpha\beta\gamma\delta}$ parameterizes the electron-electron interactions (here $\alpha, \beta, \gamma, \delta$ contain all the labels of the fermion operator). In the atomic limit, where the overlap between orbitals at different sites decays exponentially, the leading contribution of the interaction term comes from density-density coupling. In the atomic limit of the 1-band case without spin-orbit coupling, the tight-binding description becomes 
\begin{equation}
    H=-\sum_{ij}t_{ij,\sigma}a^\dagger_{i,\sigma} a_{j,\sigma}+U\sum_{i}n_{i,\uparrow}n_{i,\downarrow},
\end{equation}
where $n_{i,\sigma}=a^\dagger_{i,\sigma}a_{i,\sigma}$. For homogeneous nearest neighbour hopping $t_{ij}=t$ for adjacent sites $i,j$.
and measuring energies in units of $t$, this Hamiltonian becomes Eq. (\ref{eq:hubbard}).

In 1D, the Fermi-Hubbard model is solvable by the Bethe ansatz, meaning that by solving the Bethe equations an efficient description of the energy and ground state exists \cite{Lieb1968}. In 1D this model shows spin-charge separation, as its quasiparticles are spinons and holons. The system belongs to the universality class of the Luttinger liquid \cite{Ovchinnikov}, away from half-filling. Exactly at half-filling, the system becomes a Mott insulator, developing  a finite gap to addition of particles.
In general, at zero temperature, at least two transitions are expected for dimensions larger than one. For small fillings, the encounter probability of two particles is small, making the interaction unimportant and the system metallic (although interactions could still affect the anomalous exponents of different correlators). At half-filling there is one electron per site on average and at large enough interaction strength $U$ it is expected that the system adopts the configuration of exactly one electron per site. This state is the Mott insulator. At larger fillings the system should again become metallic, where the carriers are holes. In general, two transitions are expected as a function of density, from metallic to the Mott insulator near half-filling, and back to metallic at higher fillings. In 1D this happens at exactly half-filling for $U>0$. In higher dimensions the location of these transitions is still unresolved. The nature of the Mott transition is a matter of debate, where different mechanisms are expected to contribute, from the localization of quasiparticles\cite{Brinkman1970} to the development of magnetic instabilities \cite{Shastry1990}.

%-------------------------------------------

\section{Previous experimental implementations of quantum algorithms for the Fermi-Hubbard model and related problems}

The Fermi-Hubbard model has long been proposed as a plausible application of quantum computing. However, to our knowledge, there has been no experimental demonstration of finding the ground state of instances of the model using a quantum computer without introducing some notion of compression.

Linke et al.\cite{linke18} found the ground state of the $1\times 2$ Fermi-Hubbard model using a discretised adiabatic algorithm implemented in an ion trap experiment. Using symmetries of the system, this problem can be mapped to 2 qubits. Linke et al.\ produced two copies of the ground state and used these to measure its entanglement as measured by R\'enyi entropy, via a controlled-swap gate. The overall circuit uses 5 qubits and 31 2-qubit gates (12 to produce each copy of the ground state, and 7 for the controlled-swap).

Montanaro and Stanisic\cite{montanaro20} demonstrated the VQE algorithm for the Fermi-Hubbard model, again on a $1\times 2$ system compressed to 2 qubits and using 2 2-qubit gates, in superconducting qubit hardware. Suchsland et al.\cite{suchsland21} used symmetries to compress a 4-site Hubbard ring at half-filling to 4 qubits, and applied the VQE algorithm on a different superconducting qubit platform to find the ground state. Their variational ansatz used 3 2-qubit gates.

Outside of the VQE paradigm, a closely related work implemented a simulation of time-dynamics of the Fermi-Hubbard model\cite{GoogleFH_TDS_2020}, starting with a ground state of the noninteracting model prepared by Givens rotations, and time-evolving according to a Trotterised version of the $1\times 8$ Fermi-Hubbard Hamiltonian with occupation number 4 or 6. That work was able to demonstrate separation of spin and charge dynamics; meaningful results were obtained for circuits of 2-qubit gate depth over 400. Different error-mitigation techniques were used in that work to those employed here: averaging over different choices of qubits, Floquet calibration, and a rescaling method. Other contrasts are that the present work includes the optimisation component of VQE, considers a $2\times L_y$ system, and computes many different observables.

Beyond Fermi-Hubbard, in related work VQE has been demonstrated in the context of quantum circuits for preparing Hartree-Fock states on up to 12 qubits with high accuracy on a Google Sycamore processor\cite{GoogleVQE_2020}. These states can be efficiently prepared via Givens rotations in a similar way to the initial state used in our VQE experiment. The VQE procedure is therefore solely used to correct for errors in these Givens rotations. As Hartree-Fock states are efficiently simulable classically, algorithms creating them are excellent benchmarks of quantum computer performance, but cannot achieve exponential speedups over classical computation.

VQE has been applied to a number of other systems in quantum chemistry. The largest such experiment that we are aware of applied a hardware-efficient ansatz combined with error-mitigation techniques to find ground states of H$_2$ and LiH, using up to 6 qubits and 2-qubit gate depth 3\cite{kandala19}. VQE has also been used to demonstrate the metal-insulator transition in H$_3$, via an experiment with 3 qubits and 2 2-qubit gates\cite{smart19}. See Ref. [\onlinecite{Hempel2018}] for a survey of many other small implementations up to 2018.

Another domain where a variational approach can be used is quantum algorithms for optimisation, via the Quantum Approximate Optimisation Algorithm\cite{farhi14} (QAOA). In QAOA, one aims to find good approximate solutions to hard constraint satisfaction problems which can be expressed as finding the ground state of a classical Hamiltonian. The QAOA algorithm optimises over parametrised quantum circuits where the elementary operations are time-evolution according to the terms in this Hamiltonian, and time-evolution according to a transverse ``mixer'' term. As QAOA is solving a problem where the goal is to output a classical bit-string, rather than (for example) to output an accurate estimate of an energy, it is substantially easier to obtain meaningful results even in the presence of high levels of noise; even if the success probability is exponentially small in the problem size, this can still be distinguishable from random noise, given enough samples. The largest QAOA experiment that we are aware of used up to 23 qubits on a Google Sycamore processor and up to 3 variational layers\cite{harrigan21}.

Finally, we remark that independent work\cite{martin2021variational} has studied the ability of a low-depth variational ansatz to represent features of the Fermi-Hubbard model such as spin correlations, double occupancies, energy and ground-state fidelity (energies and fidelities were already computed for $U=2$ in Ref.\cite{cade20}). The results presented are for 1D chains on 8 sites and are obtained using classical simulation.

\section{Implementation of the Efficient Hamiltonian Variational ansatz}
\label{app:ansatz}

In this section we describe further implementation details for the variational ansatz that we used. This ansatz is based on the Hamiltonian Variational ansatz\cite{wecker15}, but with some of the hopping terms implemented using fermionic swap networks. This leads to the terms being implemented in a particular, fixed order, which can affect the performance of the quantum algorithm\cite{cade20}. We therefore refer to this ansatz specifically as the Efficient Hamiltonian Variational (EHV) ansatz.

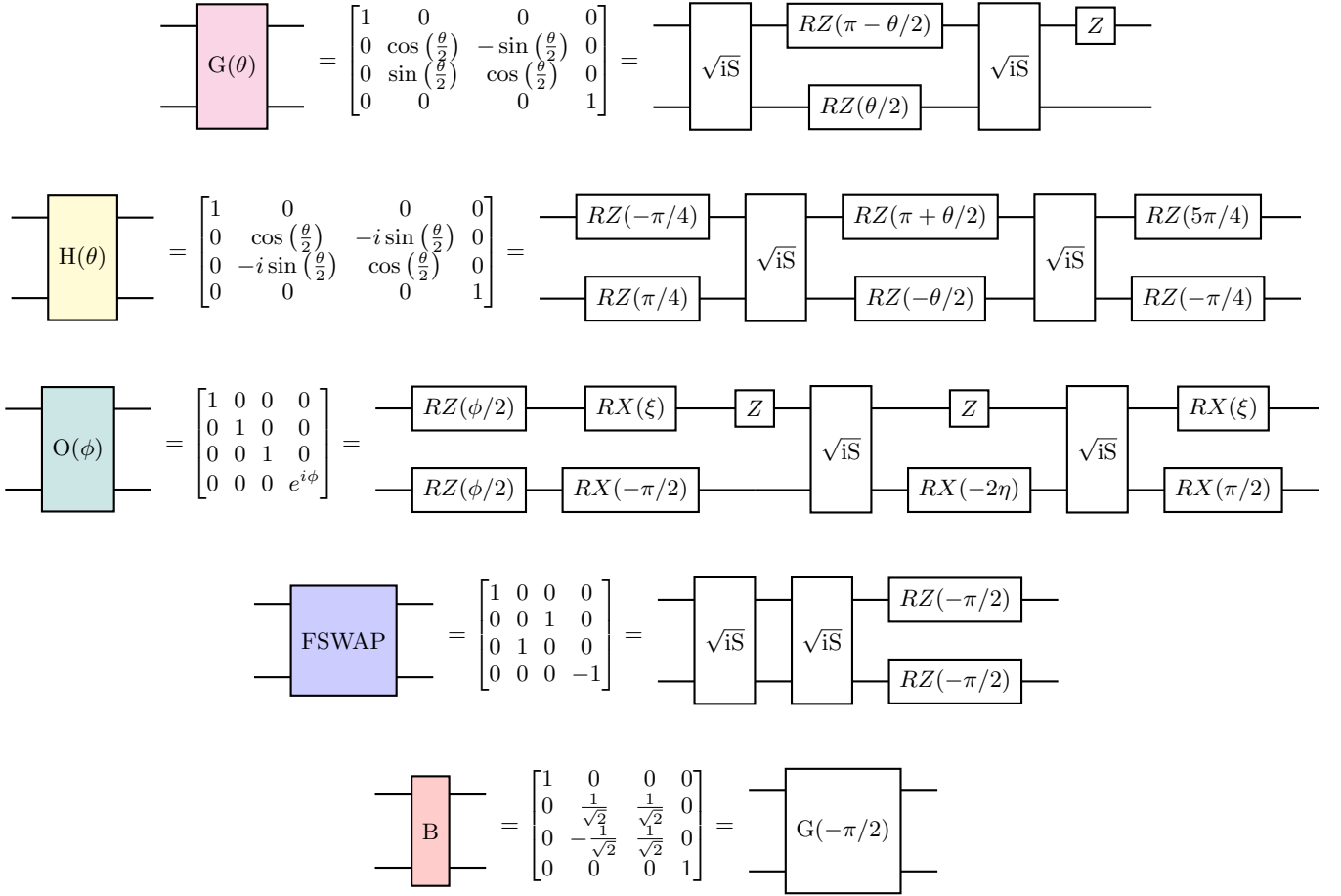
\begin{figure*}[t]
    %\begin{subfigure}[b]{\textwidth}
        \begin{equation*}
            \begin{quantikz}
            &\gate[wires=2, style={fill=magenta!20}]{\mathrm{G} (\theta)} & \qw\\
            & & \qw
            \end{quantikz} =
            \begin{bmatrix}
                1 & 0 & 0 & 0\\
                0 & \cos{\left(\frac{\theta}{2} \right)} & - \sin{\left(\frac{\theta}{2}  \right)} & 0\\
                0 & \sin{\left(\frac{\theta}{2}  \right)} & \cos{\left(\frac{\theta}{2}  \right)} & 0\\
                0 & 0 & 0 & 1
            \end{bmatrix} =
            \begin{quantikz}
            &\gate[wires=2]{\sqrt{\mathrm{iS}}}& \gate{RZ(\pi-\theta/2)} &\gate[wires=2]{\sqrt{\mathrm{iS}}}& \gate{Z} &  \qw \\
            &                                  & \gate{RZ(\theta/2)}    &                                   & \qw      &  \qw
            \end{quantikz}            
        \end{equation*} \\
    %\end{subfigure}
    %\begin{subfigure}[b]{\textwidth}
        \begin{equation*}
            \begin{quantikz}
                &\gate[wires=2, style={fill=yellow!20}]{\mathrm{H} (\theta)} & \qw\\
                && \qw
            \end{quantikz}
            = \begin{bmatrix}
                1 & 0 & 0 & 0\\
                0 & \cos{\left(\frac{\theta}{2} \right)} & - i \sin{\left(\frac{\theta}{2}  \right)} & 0\\
                0 & - i \sin{\left(\frac{\theta}{2}  \right)} & \cos{\left(\frac{\theta}{2}  \right)} & 0\\
                0 & 0 & 0 & 1
            \end{bmatrix} =
            \begin{quantikz}
            & \gate{RZ(-\pi/4)} &\gate[wires=2]{\sqrt{\mathrm{iS}}}& \gate{RZ(\pi+\theta/2)}  &\gate[wires=2]{\sqrt{\mathrm{iS}}}&  \gate{RZ(5\pi/4)}  &  \qw \\
            & \gate{RZ(\pi/4)}  &                                  & \gate{RZ(-\theta/2)} &                                 & \gate{RZ(-\pi/4)}   &  \qw
            \end{quantikz}            
        \end{equation*} \\
%    \end{subfigure}\\
%    \begin{subfigure}[b]{\textwidth}
        \centering
        \begin{equation*}
            \begin{quantikz}
            &\gate[wires=2, style={fill=teal!20}]{\mathrm{O}(\phi)} &  \qw\\
            && \qw
            \end{quantikz} =
            \begin{bmatrix}
                1 & 0 & 0 & 0\\
                0 & 1 & 0 & 0\\
                0 & 0 & 1 & 0\\
                0 & 0 & 0 & e^{i \phi}
            \end{bmatrix} =
            \begin{quantikz}
                & \gate{RZ(\phi/2)} & \gate{RX(\xi)} & \gate{Z} &\gate[wires=2]{\sqrt{\mathrm{iS}}}& \gate{Z}          &\gate[wires=2]{\sqrt{\mathrm{iS}}}& \gate{RX(\xi)}    &  \qw \\
                & \gate{RZ(\phi/2)} & \gate{RX(-\pi/2)} & \qw    &                                  & \gate{RX(-2\eta)} &                                  & \gate{RX(\pi/2)}  &  \qw
            \end{quantikz} 
        \end{equation*} \\
    %\end{subfigure}
    %\begin{subfigure}[b]{\textwidth}
        \begin{equation*}
            \begin{quantikz}
                &\gate[wires=2, style={fill=blue!20}]{\mathrm{FSWAP}}& \qw \\
                && \qw
            \end{quantikz} =
            \begin{bmatrix}
                1 & 0 & 0 & 0\\
                0 & 0 & 1 & 0\\
                0 & 1 & 0 & 0\\
                0 & 0 & 0 & -1
            \end{bmatrix} = 
            \begin{quantikz}
            &\gate[wires=2]{\sqrt{\mathrm{iS}}} &\gate[wires=2]{\sqrt{\mathrm{iS}}}&  \gate{RZ(-\pi/2)} &  \qw \\
            &                                  &                                   &  \gate{RZ(-\pi/2)}   &  \qw
            \end{quantikz}  
        \end{equation*} \\
 %   \end{subfigure}\\
 %   \begin{subfigure}[b]{\textwidth}
        \begin{equation*}
            \begin{quantikz}
                &\gate[wires=2, style={fill=red!20}]{\mathrm{B} } & \qw\\
                && \qw
            \end{quantikz} =
            \begin{bmatrix}
                1 & 0 & 0 & 0\\
                0 & \frac{1}{\sqrt{2}} & \frac{1}{\sqrt{2}} & 0\\
                0 & -\frac{1}{\sqrt{2}} & \frac{1}{\sqrt{2}} & 0\\
                0 & 0 & 0 & 1
            \end{bmatrix} =
            \begin{quantikz}
                &\gate[wires=2]{\mathrm{G} (-\pi/2)}  &  \qw\\
                &                                    &  \qw
            \end{quantikz} 
        \end{equation*} 
 %   \end{subfigure}
    \caption{Operations used within the Fermi-Hubbard VQE circuit -- Givens rotations, hopping terms, onsite terms, fermionic swaps, and basis changes for hopping term measurement -- and how they can be decomposed in terms of 1 and 2-qubit gates. Here, $\eta = \arcsin(\sqrt{2}\sin(\phi/4))$, $\xi = \arctan(\tan(\eta) / \sqrt{2})$, and $\phi \in [-\pi,\pi]$.
    }
    \label{fig:gates}
\end{figure*}

\textbf{Quantum gates.} There are five operations that we need as building blocks for our circuit, each of which is implemented using two hardware-native $\sqrt{\mathrm{iSWAP}}$ gates and some single-qubit gates.
The initial state is prepared using Givens rotations (gate $\mathrm{G}$ in Fig.~\ref{fig:gates}).
Then each layer of the EHV ansatz consists of onsite (gate $\mathrm{O}$ in Fig.~\ref{fig:gates}) and hopping (gate $\mathrm{H}$ in Fig.~\ref{fig:gates}) gates, corresponding to time-evolution by onsite and hopping terms in the Fermi-Hubbard Hamiltonian (\ref{eq:hubbard}), respectively, where hopping terms are assumed to act only on adjacent modes in the Jordan-Wigner transform. We have
\[ \mathrm{H}(\theta) = e^{-i\theta(XX+YY)/2},\;\;\;\; \mathrm{O}(\phi) = e^{i\phi \proj{11}}. \]
For a lattice with shape $2 \times L_y$, we need a fermionic SWAP (FSWAP) gate to implement the fermionic swap network (gate $\mathrm{FSWAP}$ in Fig.~\ref{fig:gates}).
Finally, we need a gate for the change of basis needed to measure the hopping terms (gate $\mathrm{B}$ in Fig.~\ref{fig:gates}). In previous work it was suggested to use a Hadamard gate within the $\{\ket{01},\ket{10}\}$ subspace\cite{cade20}; here we use an equivalent operation that can be implemented more easily. Note that this operation preserves occupation number, which allows the use of error detection.

When implemented on hardware, the single-qubit gates shown in Fig.~\ref{fig:gates} are decomposed in terms of the hardware-native PhasedXZ gate primitive. Due to a sign error in our implementation of this decomposition, in the experiments the onsite gate $\mathrm{O}(\phi)$ was implemented up to identical single-qubit Z rotations on each qubit, which leave the overall state unchanged within a fixed occupation number subspace. Spot checks comparing with a correctly decomposed onsite gate confirmed that, as expected, these Z rotations did not affect the overall accuracy of the experiment.

\textbf{State preparation.} The first step of the EHV ansatz is to prepare the ground state of the noninteracting ($U=0$) Fermi-Hubbard Hamiltonian. Preparation of this state has been studied extensively before and an efficient algorithm using Givens rotations is known\cite{jiang2018quantum} which achieves circuit depth $N-1$, and a total of $(N - \Noccsigma) \Noccsigma$ Givens rotations (for each spin sector), where $N$ is the number of modes per spin sector, or equivalently the size $L$ of the lattice ($L=L_x \times L_y$) and $\Noccsigma$ is the number of fermions in the spin sector $\sigma$. A detailed analysis of alternative state preparation methods\cite{cade20} concluded that this algorithm was the most efficient known for small system sizes.
To prepare this initial state we use the OpenFermion\cite{mcclean20} implementation of this algorithm.

\textbf{Swap operations.} There are two types of swap operations that are needed in our algorithm: fermionic swaps (FSWAPs) to rearrange the Jordan-Wigner ordering, and physical (standard) swaps to bring distant qubits together. An FSWAP operation can be implemented with two native gates, as shown in Fig.~\ref{fig:gates}, whereas physical swaps would require three native gates. However, in our experiment we are always able to use fermionic swaps in place of physical swaps. The one place where physical swaps would naturally be used is rearranging qubits before and after implementing an onsite (CPHASE) gate. As the onsite gates are diagonal, the sign part of the FSWAP gates commutes with them and cancels out.

\textbf{Measurement.} Measuring the energy of the VQE state can be achieved with 3 different measurement circuits for $1 \times L_y$ instances (vertical hopping 1 and 2, and onsite), and with 4 circuits for $2\times L_y$ instances (horizontal hopping, vertical hopping 1 and 2, and onsite). Onsite energy is measured via a computational basis measurement and counting the number of sites where both spin-up and spin-down qubits receive a 1 outcome. For $1 \times L_y$, each type of vertical hopping term is measured using a layer of basis transformations, using the $B$ gate shown in Fig.\ \ref{fig:gates}. These gates diagonalise the hopping terms, enabling the corresponding energy to be measured via a computational basis measurement. The second type of vertical hopping measurement can be merged into the final layer of gates in the circuit (Fig.\ \ref{fig:2x4_swap}) to reduce the quantum circuit depth. Measuring the energy for $2 \times L_y$ instances is similar, except that vertical hopping terms are split up in a different way (also see Fig.\ \ref{fig:2x4_swap}), and some of them require a layer of fermionic swap gates before measurement.

\textbf{Quantum circuit complexity.} The complexities of the largest circuits that we executed are summarised in Table \ref{tab:complexities}. It is interesting to note that for a $2\times 4$ lattice, the most complex instances in terms of circuit complexity were not at half-filling; this is because in this specific case, one fewer layer of Givens rotations is present than worst-case bounds would suggest\cite{jiang2018quantum}.

\textbf{Scaling of algorithms using rectangular and zig-zag configurations.} The quantum circuit depth for each layer of the EHV ansatz for a $L_x \times L_y$ lattice, $L_x \le L_y$, with no restrictions on quantum circuit topology\cite{cade20}, and assuming that an arbitrary 2-qubit operation can be implemented with one hardware-native gate, is as low as $2L_x+1$ (for even $L_x$). Almost all gates that occur in the algorithm act across nearest neighbours in the Jordan-Wigner line, with the exception of onsite gates and basis transformations necessary for measuring vertical hopping terms.

To implement this circuit using a $L_xL_y \times 2$ rectangular configuration of qubits on a device, we can associate one row with each spin-type, and associate modes within each spin sector with qubits in the Jordan-Wigner ordering (see Ref. [\onlinecite{cai20}] for a related proposal). Then onsite gates are local, so there is no additional cost per ansatz layer, and the only remaining long-range transformation is the basis transformations for measuring vertical hopping terms.

To implement these, it is sufficient to reorder a pair of rows in the Fermi-Hubbard lattice such that vertically neighbouring pairs become horizontally neighbouring. It is easiest to illustrate a procedure for this with an example. If we label modes in the first row of a $4 \times L_y$ lattice as $1,2,3,4$, and modes in the second row as $A,B,C,D$, we want to transform from the ordering $1234ABCD$ to the ordering $1D2C3B4A$. This transformation can be split into two parts: flipping $ABCD$ to $DCBA$, and then transforming $1234DCBA$ to $1D2C3B4A$. For rows of length $L_x$ the first part can be implemented with $L_x$ layers of FSWAP operations in an alternating even-odd pattern\cite{kivlichan18}. The second part requires $L_x-1$ layers of FSWAPs, in a triangle configuration beginning at the middle, to interleave the first and second rows. The overall cost is therefore $2L_x-1$ layers of FSWAP operations; for the special case of a $2\times L_y$ lattice, we can improve this to just 1 layer (Fig. \ref{fig:2x4_swap}).

All of the same arguments hold for the zig-zag configuration that we use in our experiments, except that now we need an additional layer of FSWAP gates before and after the onsite gates, giving an overall cost of $2L_x+3$ 2-qubit gate depth per ansatz layer. We remark that previous work giving complexity bounds for nearest-neighbour and Sycamore architectures\cite{cade20} considered more ``balanced'' configurations suitable for fitting more modes into a small quantum processor whose width and height are closer in size; this led to larger bounds ($4L_x+1$ for nearest-neighbour and $6L_x+1$ for Sycamore, respectively).

\begin{table}[]
    \centering
    \begin{tabular}{|c|c|c|c|c|c|}
    \hline
        Lattice & $\Nocc$ & Embedding & \begin{tabular}[c]{@{}l@{}}Circuit\\ depth\end{tabular} & \begin{tabular}[c]{@{}l@{}}2-qubit\\ depth\end{tabular} &
        2-qubit gates\\
        \hline $1\times 4$ & 4 & Rectangle & 41 & 20 & 64\\
        \hline $1\times 8$ & 8 & Zig-zag & 53 & 26 & 140\\
        \hline $2\times 4$ & 7 & Zig-zag & 65 & 32 & 176\\
        \hline
    \end{tabular}
    \caption{Largest circuit complexities for the configurations considered in our experiments. The $1\times 4$ experiments used a depth 2 VQE ansatz, while the other experiments used a depth 1 ansatz. Circuit complexities depend on which energy measurement is being performed in the VQE algorithm; stated complexities are the maximal ones over these circuits, showing the occupation numbers where these are  achieved.}
    \label{tab:complexities}
\end{table}

%-------------------------------------------

\section{Variational optimiser}
\label{sec:variational-optimiser}

In this work we introduce a new variational optimisation method, which we call \emph{Bayesian model gradient descent (BayesMGD)}, and compare it with the standard simultaneous perturbation stochastic
approximation (SPSA) algorithm\cite{spsa}, which has been previously successfully used as an optimization algorithm for VQE on
superconducting quantum computers \cite{kandala17,Ganzhorn2019}, and the model gradient descent (MGD) algorithm introduced by Sung et al.\cite{sung20} for 
precisely the task of optimising parametric quantum circuits\cite{harrigan21}.

The main idea
of MGD is to sample points and function values $(\vec\theta_i, y_i)$ in 
a trust region around $\vec\theta$, fit a quadratic surrogate model using 
linear least squares to all data available in the trust region and use this
surrogate model to estimate the gradient. Our algorithm is designed to improve on these
ideas via Bayesian analysis. We perform iterative, Bayesian updates on the surrogate 
model and utilise the sample variance to estimate the uncertainty in the 
fit parameters and surrogate model evaluations. 
Utilising the sample variance to estimate the uncertainty of function evaluations allows
for more accurate surrogate models and estimating the uncertainty in the surrogate 
model evaluations allows us to put error bars on the predictions.

We are given a random field $f(\vec\theta)$ (that is, a collection of random variables parameterised by $\vec\theta$) and want to find the parameters 
$\vec\theta$ such that the expectation value
$\mu(\vec\theta) := \mathds{E}[f(\vec\theta)]$ is minimal. We assume that
at each $\vec\theta$ the variance of the random variable $f(\vec\theta)$ is finite, such that the central limit theorem is applicable to sample means of $f(\vec\theta)$.
Since we are always interested in situations where we take many samples at a given 
$\vec\theta$ and approximate $\mu(\vec\theta)$ by their mean, we can equivalently assume
--- and will from now on ---
that $f(\vec\theta)$ is normally distributed at each $\vec\theta$ with known variance
$\sigma^2(\vec\theta)$.
Furthermore, the mean function $\mu(\vec\theta)$ is assumed to be smooth and hence 
it is locally always well described by a quadratic surrogate model 
\begin{equation}
  f_s(\vec\theta; \vec\beta) = \beta_0
                             + \sum_{j=1}^{n_c} \beta_j \theta_j
                             % + \sum_{j<k}^n \beta_{jk} \theta_j \theta_k
                             + \sum_{j,k=1,~j<k}^{n_c} \beta_{jk} \theta_j \theta_k
  \label{eq:surrogate_model}
\end{equation}
which is linear in its model parameters $\beta_0, \beta_j$ and $\beta_{jk}$, and where $n_c$ is the number of circuit parameters.

In each iteration $m$ we randomly pick $p = \eta\, \mathrm{dim}(\beta)$ sampling points
$\vec\theta^{(i)}$ in a $\delta_m$-ball around $\vec\theta_m$ and get noisy function
evaluations $y_i \sim \mathcal{N}(\mu(\vec\theta^{(i)}), \sigma_i^2)$ with
\emph{approximately known} uncertainty $\sigma_i$, where
$\eta$ is the ratio between the number of new sampling points $p$ and the
number of points needed for a fully determined quadratic fit.
The sampling radius scales as $\delta_m = \delta / m^\xi$ with a sample radius
decay exponent $\xi$ and initial sampling radius $\delta$. This
new data $\{\vec\theta^{(i)}\}$ and $\{y_i, \sigma_i\}$ is used to update our
belief $p_{m|m-1}(\vec\beta)$ about the parameters $\vec\beta$ at the $m$-th step given
the data up until step $m-1$  using Bayes' rule to a new belief $p_{m|m}(\vec\beta)$ incorporating the new data from the $m$-th step: 
\begin{equation}
\begin{aligned}
  p_{m|m}(\vec\beta) =& P(\vec\beta \,|\, \{\vec\theta^{(i)}\}, \{y_i, \sigma_i\}) \\
                 \propto& P(\{y_i, \sigma_i\}\,|\, \{\vec\theta^{(i)}\}, \vec\beta)
                  \, p_{m|m-1}(\vec\beta) \\ 
                 =& \prod_{i=1}^p \mathcal{N}(y_i; f_s(\vec\theta^{(i)}; \vec\beta), \sigma_i)  
                    \\ & \times 
                        \mathcal{N}(\vec\beta ; \vec\beta_{m|m-1}, \vec\Sigma_{m|m-1}) \\
                 =:& \mathcal{N}(\vec\beta ; \vec\beta_{m|m}, \vec\Sigma_{m|m}),
\end{aligned}
\label{eq:bayesmgd-update}
\end{equation}
where in the last line we use the fact that the product of Gaussians is again
a Gaussian to implicitly define $\vec\beta_{m|m}$ and $\vec\Sigma_{m|m}$. We defer the detailed 
derivation of $\vec\beta_{m|m}, \vec\Sigma_{m|m}$ in terms of the prior
$\vec\beta_{m|m-1}, \vec\Sigma_{m|m-1}$ and new data to Section
\ref{sec:bayesmgd-kalman-filters}, together with a discussion of the relation of 
BayesMGD and Kalman filters and pseudo-code for the algorithm.

Since the surrogate model $f_s(\vec\theta; \vec\beta)$ is linear in the model parameters
$\vec\beta$ the usual uncertainty propagation formulas are exact and we know that
\begin{equation}
  f_s(\vec\theta_m; \vec\beta) \sim \mathcal{N}
  \left(
    f_s(\vec\theta_m; \vec\beta_{m|m}), (\nabla_{\vec\beta} f_s)^\dagger \vec\Sigma_{m|m} \nabla_{\vec\beta} f_s
  \right),
\end{equation}
where $\nabla_{\vec\beta} f_s$ denotes the gradient of $f_s$ with respect to $\vec\beta$
evaluated at $(\vec\theta_m; \vec\beta_{m|m})$. Similarly, we also obtain a distribution
over the gradient $\nabla_{\vec\theta} f_s(\vec\theta_m; \vec\beta)$. 
The maximum a posteriori estimate for the gradient is simply obtained by plugging 
the most likely value $\vec\beta_{m}$ for the model parameters $\vec\beta$ 
into the gradient of the surrogate model:
\begin{equation}
  g(\vec\theta_m) = \nabla_{\vec\theta} f_s(\vec\theta; \vec\beta_{m|m}).
\end{equation}
With this estimate of the gradient we perform a gradient descent step 
\begin{equation}
  \vec\theta_{m+1} = \vec\theta_m - \gamma_m g(\vec\theta_m).
\end{equation}
Here, $\gamma_m = \gamma / (m + A)^\alpha$ is the gradient step
width with a stability constant $A$, decay exponent $\alpha$ and initial step width
$\gamma$.

Changing $\vec\theta$ does not change the local surrogate model, but it adds uncertainty proportional to the step width to it. Hence the belief at $\vec\theta_{m+1}$ without
data at that point is described by
\begin{equation}
\begin{aligned}
  \vec\beta_{m+1|m} &= \vec\beta_{m|m}  \\
  \vec\Sigma_{m+1|m} &= \vec\Sigma_{m|m} + \frac{\gamma_m^2 |g(\vec\theta_m)|^2}{l^2} \mathds{1},
\end{aligned}
\end{equation}
where $l$ is the length scale on which our
quadratic model becomes invalid.  The choice of adding uncertainty proportional
to the squared step width is heuristic so far, but can be motivated using
Gaussian processes.
A Gaussian process is a probability distribution over functions that allows one, among other things, to compute the probability distribution of function
value, gradient and hessian at some point $\vec\theta_{m+1}$ conditioned
on the function value, gradient, and hessian available at some previous
point $\vec\theta_m$.  For a Gaussian process with a squared exponential kernel and
small $|\vec\theta_m - \vec\theta_{m+1}|$ the uncertainty about function value, gradient,
and hessian at $\vec\theta_{m+1}$ grows with the squared distance from $\vec\theta_m$.
The exact rate at which the uncertainty for each of the entries of $\vec\beta_m$ 
grows requires in-depth analysis that we replaced with uniform scaling in all
entries.

\emph{Note added ---} In independent recent work\cite{tamiya2021stochastic} 
another optimisation algorithm, called SGSLBO (Stochastic Gradient Line Bayesian Optimization), is proposed for VQE that is at first glance similar to ours. However, this algorithm is based on the use of stochastic gradient descent to determine the gradient direction paired with Bayesian optimisation for a line search along the gradient direction. In our case ``Bayesian'' refers to the iterative Bayesian procedure we use to update the model parameters $\vec{\beta}$.

%-------------------------------------------

\subsection{Details of implementation parameters}

\begin{table}[t]
    \centering
    \begin{tabular}{|c|c|c|c|c|c|}
    \hline
        Instance & Params & Points/iter & Max evals & Max iters\\
        \hline
        $1\times 4$ & 6 & 42 & 2520 & 30 \\
        $1\times 8$ & 3 & 15 & 300 & 10 \\
        $2\times 4$ & 4 & 23 & 600 & 14 \\
        \hline 
    \end{tabular}
    \caption{BayesMGD characteristics in experiments. $1\times 4$ instances had two variational layers, others had one variational layer.}
    \label{tab:bayesmgd}
\end{table}

Characterising the VQE ground state for a given Fermi-Hubbard instance can be separated into two parts: the VQE part, which runs the BayesMGD algorithm to determine the optimal variational parameters for the quantum circuit; and the state preparation part, which uses these parameters to produce copies of the VQE ground state itself, and also many other FLO states used for error mitigation (see Appendix \ref{app:tflo}). These parts can be carried out at different times, which may be advantageous, as device performance fluctuates over time. The state preparation part uses all error mitigation techniques described in Appendix \ref{app:error_mitigation}, whereas for efficiency the VQE part does not use particle-hole symmetry or TFLO.

\textbf{VQE part.} In all cases, the BayesMGD optimiser used 1000 shots (energy measurements) per evaluation point, multiplied by 2 for evaluating at the parameters and their negations (see Appendix \ref{app:symmetries}). Bounds on numbers of evaluations are shown in Table \ref{tab:bayesmgd}. For all instances, hyperparameters $\eta=1.5$, $\delta = 0.6$, $\xi = 0.101$, $l = 0.2$ were used.
For $1\times 4$ and $1\times 8$, $\gamma = 0.3$, $A=1$ were used, whereas for $2\times 4$, $\gamma = 0.6$, $A=2$ were used. Increasing $\gamma$ moves through the parameter space more aggressively, and increasing the stability parameter $A$ reduces the chance of overaggressive moves at the start of the algorithm. Wall clock time for completing a VQE run was under 30 minutes for $1\times 8$ instances, and under 70 minutes for $2\times 4$ instances. We split the circuits evaluated into batches of size at most 80 to avoid timeout and circuit size constraints imposed by the quantum cloud platform.

\textbf{State preparation part.} In this part, we compute the energy of the VQE ground state by taking the average over 100,000 energy measurements, again both at the VQE parameter values and their negations. In order to use TFLO, we also evaluate the energy at the closest FLO point (the one where the onsite parameters are set to 0), with 100,000 energy measurements; and also at 16 other points (and their negations), which have been chosen such that their exact energies are well-spaced. For each of these 16 points we perform 20,000 energy measurements. We carried out this procedure 3 times for each instance. Wall clock times are up to approximately 8 minutes per run for $1\times 8$ and $2\times 4$ instances.

%-------------------------------------------

\section{Error mitigation}
\label{app:error_mitigation}

The error mitigation techniques we used can be divided into three categories: low-level techniques which are not specific to the Fermi-Hubbard model and are targeted at the particular hardware platform used; techniques based on the symmetries of the Fermi-Hubbard model; and a technique that is designed to mitigate errors in fermionic simulation algorithms. For a survey of other error mitigation techniques, see Ref. [\onlinecite{endo20}].

%-------------------------------------------

\subsection{Low-level error mitigation}

\textbf{Circuit structure.} We ensure that our quantum circuits are of a form where we alternate between layers of single-qubit and two-qubit gates. This is similar to the circuit topology used for quantum supremacy experiments\cite{sycamore}, and is advantageous because two-qubit gates generally take longer to execute than single-qubit gates, and the time taken to execute a layer is equal to the time for the slowest gate in it.

\textbf{Qubit selection.} Our experiments use a set of up to 16 qubits in a particular orientation. On the quantum processor we used, there were four such sets of 16 qubits available, and the choice of which set to use could make a significant difference to experimental performance. In a VQE experiment, a straightforward metric to use to select qubits is the energy measured for some choice of parameters. In general, measuring lower energies is better, as we expect decoherence to increase the energy; coherent errors will also usually increase the energy, if the initial state's energy is close to the lowest possible within that VQE ansatz. We therefore selected qubits for our experimental run by choosing fixed parameters (all zeroes), measuring the energy at these parameters on all four sets of possible qubits, and choosing the qubits which achieved the lowest energy. Spot checks throughout the experimental run showed that this set of qubits remained high-quality. Unlike some previous work\cite{GoogleFH_TDS_2020}, we found that averaging results over different qubit sets was not advantageous in reducing the final error (even when other error-mitigation techniques were also applied); it was usually better just to pick one ``best'' set of qubits and use them throughout.

\textbf{Run selection.} For each set of VQE parameters, we carried out 3 experiments at different times to create the corresponding VQE state and generate energy estimates. We then selected the run which returned the lowest ``raw'' energy estimate (following postselection by occupation number (see below)) for subsequent calculations. The intent behind this is that we expect the level of noise and errors experienced by the qubits at a particular time to be correlated with the measured energy, so a lower energy should correspond to a higher-accuracy experiment.

\begin{figure}
    \centering
    \begin{minipage}[c]{\linewidth}
        \subcaption{(a) Structure of circuit with X layers}
        \includegraphics[width=0.95\linewidth]{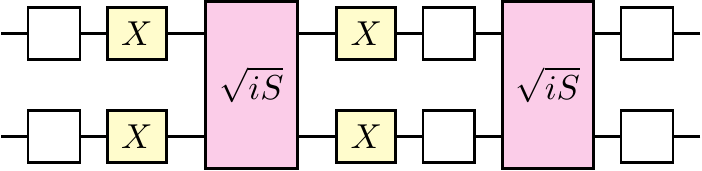}
    \end{minipage}
    %\vspace{11pt}
    
    \vspace{10pt}
    
    \begin{minipage}[c]{\linewidth}
        \subcaption{(b) Stability over time with and without X layers}
        \includegraphics[width=0.95\linewidth]{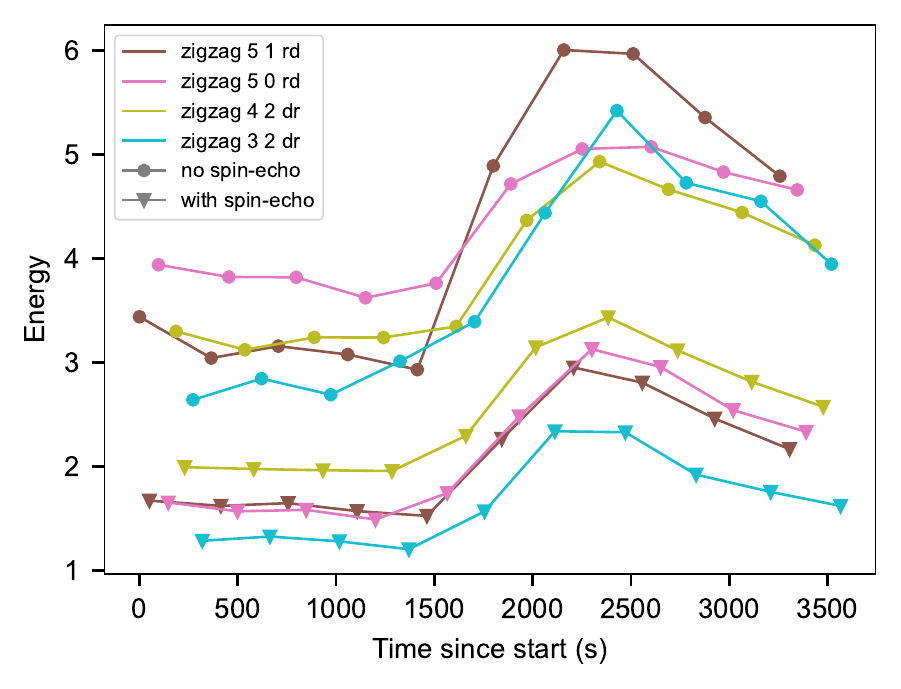}
    \end{minipage}
    \caption{Spin-echo technique for reducing errors. Graph shows the energy obtained from measuring at the lowest-energy parameters achievable in VQE depth 1 for a $1\times 8$ instance at half-filling. Lines with triangle markers are with spin-echo, lines with circle markers are without. Exact energy is $-3.478$ so lower energies are better. 4 sets of qubits were tested in zigzag configurations; labels indicate starting positions and direction (`rd': right then down; `dr': down then right)}.
    \label{fig:spinecho}
\end{figure}

\textbf{Spin echo.}  We used a technique inspired by the concept of refocusing by spin echo in NMR\cite{freeman98}. Given an unknown unitary operation $U = e^{i \theta Z}$, then as $XUX = U^{-1}$, implementing the sequence $UXUX$ produces the identity map. If we think of $U$ as an error whose precise form is unknown, this allows the error to be corrected. Here we implemented this idea by sandwiching alternating layers of 2-qubit gates with layers of X gates (Fig.\ \ref{fig:spinecho}). As $X^{\otimes 2}$ commutes with the $\sqrt{\mathrm{iSWAP}}$ gate, on a perfect quantum computer this would leave the unitary operation implemented unchanged. On an imperfect quantum computer, it may help to correct errors. Introduction of X gates was previously found to reduce errors on an idle qubit in a similar superconducting quantum processor\cite{Google_2021_topological}, and more complex dynamical decoupling sequences\cite{viola98} have been demonstrated to significantly reduce decoherence in other superconducting quantum computing systems\cite{pokharel18}.

In our experiments the effect on errors was substantial, whichever subset of qubits was used. As well as correcting unwanted Z rotations, another possible explanation for this effect may be that 2-qubit gates are known to experience a substantial ``parasitic CPHASE\cite{GoogleFH_TDS_2020}'', manifesting as an undesired phase acting on $\ket{11}$. The X gates move this phase to $\ket{00}$, which may reduce its effect over the circuit as a whole. Also observe from Fig.\ \ref{fig:spinecho} the substantial variations in measured energies over time.

%-------------------------------------------

\subsection{Symmetries of the Fermi-Hubbard model}
\label{app:symmetries}

The Fermi-Hubbard model has a number of symmetries which can be used to mitigate errors: time-reversal symmetry, particle number conservation per spin sector, particle-hole symmetry, and reflection symmetry of the lattice.

\textbf{Time-reversal symmetry} of the terms in the Fermi-Hubbard model implies that the VQE energy is unchanged if all parameters are negated\footnote{Technically, this is a consequence of time-reversal symmetry combined with spin-flip symmetry.}. This is because the initial state $\ket{\psi_0}$, the ground state of the noninteracting Fermi-Hubbard model, is real in the computational basis, and each of the interaction terms $H_k$ in the Fermi-Hubbard Hamiltonian is symmetric. Hence
\[ e^{-i \theta_k H_k} = \left(e^{i \theta_k H_k}\right)^\dag = \left(e^{i \theta_k H_k}\right)^*, \]
where $*$ denotes complex conjugation in the computational basis. The final state of the algorithm for parameters $\{-\theta_k\}$ is then
\begin{eqnarray*}
\ket{\psi_f} &=& \prod_k e^{-i \theta_k H_k} \ket{\psi_0} = \prod_k (e^{i \theta_k H_k})^* \ket{\psi_0}\\
&=& \left( \prod_k e^{i \theta_k H_k} \ket{\psi_0}\right)^*.
\end{eqnarray*}
The energy of this state with respect to the overall Fermi-Hubbard Hamiltonian $H$ is then equal to that achieved by the final state with parameters $\{\theta_k\}$, because
\[ \braket{\psi_f|H|\psi_f} = \braket{\psi_f|^*H|\psi_f}^*. \]
We can take advantage of this symmetry by evaluating the energy for a given set of VQE parameters $\vec{\theta}$ as the average of the energies at $\vec{\theta}$ and $-\vec{\theta}$. The intent is that this will mitigate the effect of systematic coherent errors such as over-rotations.

\textbf{Particle number conservation per spin sector} in the Fermi-Hubbard model follows from invariance of the Fermi-Hubbard Hamiltonian under the $U(1)\times U(1)$ symmetry $a^\dag_{i,\sigma} \mapsto e^{i\alpha_{\sigma}} a^\dag_{i,\sigma}$ (i.e.\ a $U(1)$ transformation acting on spin-up and spin-down modes independently). As a consequence, the overall Fermi-Hubbard Hamiltonian preserves the occupation number (Hamming weight) within spin-up and spin-down sectors.

Further, occupation number preservation within each spin sector holds for all operations in the Hamiltonian variational ansatz. As the quantum algorithm starts with a state with a known occupation number in each spin sector, this enables us to reject any runs with an incorrect final occupation number. In particular, this detects many errors that occur due to incorrect qubit readout, which can be a significant source of error in superconducting qubit systems (for example, realistic estimates could be a 1\% probability of a 0 being incorrectly read as a 1, and a 5\% probability of a 1 being incorrectly read as a 0). We found that in our 16-qubit experiments, between 7\% and 29\% of runs were retained due to having correct occupation numbers (see Table \ref{tab:postselection}).

It is interesting to note that we expect that checking the occupation number within each spin sector should be sufficient to detect the vast majority of readout errors, without the need for additional readout error mitigation techniques\cite{endo18,kandala17,maciejewski19,chen2019}. A rough upper bound on the probability that there is an undetected readout error can be found by multiplying the probability that there is a pair of errors $0 \mapsto 1$ and $1 \mapsto 0$ by the number of pairs of qubits within each spin sector, and then by 2 for the number of spin types. Assuming independent readout errors, the probability of such a pair of errors can be roughly upper-bounded by $10^{-3}$, and in a 16-qubit experiment we have 8 qubits in each spin sector, giving $\binom{8}{2} = 28$ possible pairs of readout errors in each spin sector, and hence an overall upper bound of less than 6\% on the probability that there is a undetected readout error.

\begin{table}[]
    \centering
    \begin{tabular}{|c|c|c|}
        \hline Lattice & Min probability & Max probability \\
        \hline $1\times 4$ & 0.32 & 0.66 \\
        $1\times 8$ & 0.12 & 0.29 \\
        $2\times 4$ & 0.077 & 0.25\\
         \hline
    \end{tabular}
    \caption{Probabilities of successfully postselecting on occupation number for the different lattices considered in our experiments.}
    \label{tab:postselection}
\end{table}

\textbf{Particle-hole symmetry} relates low to high fillings in the Fermi-Hubbard model.
On a bipartite lattice we define two sublattices $A$ and $B$, such that each neighbour of a site in $A$ belongs to $B$ and vice versa. The particle-hole transformation $\mathcal{P}$ acts as $\mathcal{P}a_{j\sigma}\mathcal{P}^{\dagger}=(-1)^{b_{j}}a_{j\sigma}^{\dagger}$, where $b_{j}=0$ if $j\in A$ and $b_{j}=1$ if $j\in B$. Under this transformation the Hamiltonian (\ref{eq:hubbard}) becomes
\begin{equation}
\mathcal{P}H\mathcal{P}^{\dagger}	=H+U(L-\Nocc)
\end{equation}
where $\Nocc=N_{\uparrow}+N_{\downarrow}$ is the number of electrons in the system and $L$ is the size of the system. The density operator $n_{i\sigma}=a_{i\sigma}^{\dagger}a_{i\sigma}$ transforms as $\mathcal{P}n_{i\sigma}\mathcal{P}^{\dagger}=a_{i\sigma}a_{i\sigma}^{\dagger}=1-n_{i\sigma}$. For a unique ground state $|GS_{(N_{\uparrow},N_{\downarrow})}\rangle$ with $N_\sigma$ electrons of spin $\sigma$, $\mathcal{P}|GS_{(N_{\uparrow},N_{\downarrow})}\rangle=|GS_{(L-N_{\uparrow},L-N_{\downarrow})}\rangle$ and the density correlations satisfy
\begin{eqnarray}\nonumber\label{eq:PH_corr}
&\langle n_{i\sigma}n_{j\sigma'}\rangle_{N_{\uparrow},N_{\downarrow}}=\langle n_{i\sigma}n_{j\sigma'}\rangle_{L-N_{\uparrow},L-N_{\downarrow}}\\
&-\langle n_{i\sigma}\rangle_{L-N_{\uparrow},L-N_{\downarrow}}-\langle n_{j\sigma'}\rangle_{L-N_{\uparrow},L-N_{\downarrow}}+1,
\end{eqnarray}
where we defined
$\langle GS_{(N_{\uparrow},N_{\downarrow})}|\mathcal{O}|GS_{(N_{\uparrow},N_{\downarrow})}\rangle =: \langle\mathcal{O}\rangle_{N_{\uparrow},N_{\downarrow}}$. The spin correlations can be obtained from ({\ref{eq:PH_corr}).

Importantly, all the terms in the Hamiltonian Variational ansatz are also essentially invariant under this symmetry. This can be seen concretely in the Jordan-Wigner transform, where the particle-hole symmetry corresponds to an $X$ gate acting on each qubit. Hopping terms in the ansatz commute with $X^{\otimes 2}$, while onsite terms (CPHASE gates) commute up to a $Z$ rotation on both qubits. As the same parameter is used for all such rotations within one layer, this becomes an unobservable global phase within each occupation number subspace. Thus $X^{\otimes N}$ effectively commutes with the entire variational circuit $C$, implying that for any observable $\mathcal{O}$,
\be \braket{\psi|X^{\otimes N} C^\dag \mathcal{O} C X^{\otimes N}|\psi} = \braket{\psi|C^\dag X^{\otimes N} \mathcal{O} X^{\otimes N} C|\psi}, \ee
and hence that we can interpret any observable on a VQE state $\ket{\psi}$ in terms of a related observable on the particle-hole transformed state $X^{\otimes N}\ket{\psi}$.

Particle-hole symmetry can thus be used for error mitigation, by producing an estimate of an observable for the VQE ground state at occupation number $\Nocc$ as an average of the experimentally obtained value at $\Nocc$ and the value at $L-\Nocc$ (suitably transformed). This effectively replaces the worst-case and best-case errors of this pair with their average.

\textbf{Reflection symmetry.} The site-dependent observables we measure (charge and spin, and the corresponding correlations) are symmetric about reflections of the lattice in the $x$ and $y$ directions. We can obtain a further reduction in worst-case error by averaging these quantities over reflections of the lattice (2 points in the case of a 1D lattice, and 4 points in the case of a 2D lattice).

%-------------------------------------------

\subsection{Training with fermionic linear optics}
\label{app:tflo}

Training with fermionic linear optics (TFLO) is a method proposed by two of us\cite{montanaro21} to mitigate errors in quantum algorithms for simulating fermionic systems, which fits into an overall framework initially introduced by Czarnik et al.\cite{czarnik20}. The idea is based on producing a set of pairs of noisy and exact energies, which are then used as training data to infer a map from the noisy energy evaluation for the final state produced by VQE to an approximation of the exact energy.

To implement this concept requires a family of quantum circuits which well-represent the error behaviour of the quantum circuit which we actually wish to evaluate. In Ref. [\onlinecite{czarnik20}], the family of circuits used was Clifford circuits, which can be simulated efficiently classically via the Gottesman-Knill theorem. Here, we use fermionic linear optics (FLO) circuits, which can also be simulated efficiently classically\cite{terhal02}. This family of circuits is tailor-made for mitigating errors in VQE for the Fermi-Hubbard model. An FLO circuit starts with a computational basis state, which corresponds to the state produced by applying some creation operators to the vacuum, and contains operations corresponding to time-evolution by quadratic Hamiltonians, via unitary operators of the form $U = e^{iH}$, where $H = \sum_{j,k} h_{jk} a_j^\dag a_k$.

In the case of VQE with the Hamiltonian variational ansatz\cite{wecker15}, all operations in the circuit are either time-evolution by terms in the Fermi-Hubbard Hamiltonian, or preparation of the initial ground state of the noninteracting Fermi-Hubbard model via Givens rotations (which are FLO). Thus almost all operations in the circuit are FLO, except for time-evolution by the onsite terms. Therefore, any VQE circuit where the onsite parameters are set to 0 is an FLO circuit and can be simulated efficiently to benchmark the behaviour of errors in the circuit.

The TFLO method has been successfully applied to VQE for a $2\times 3$ instance of the Fermi-Hubbard model in classical emulation (with a simple depolarising noise model), and to a $1\times 2$ instance on real quantum hardware\cite{montanaro21}, reducing errors by a factor of 10--30 or more. However, as with other error mitigation techniques, it is unclear in advance how well TFLO will perform in a given experiment, especially for larger instance sizes.

\begin{figure*}
    \centering
    \begin{minipage}[c]{0.32\textwidth}
        \subcaption{(a) $1\times8$, $U=8$, depth 1}
        \includegraphics[scale=0.95]{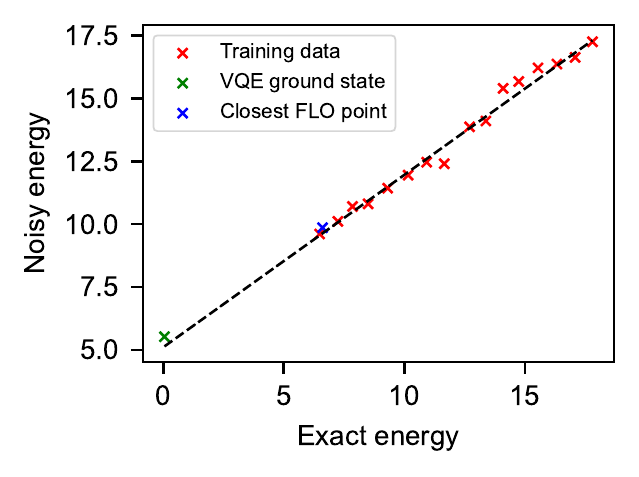}
    \end{minipage}
    \hfill
    \begin{minipage}[c]{0.32\textwidth}
        \subcaption{(b) $2\times4$, $U=4$, depth 1}
        \includegraphics[scale=0.95]{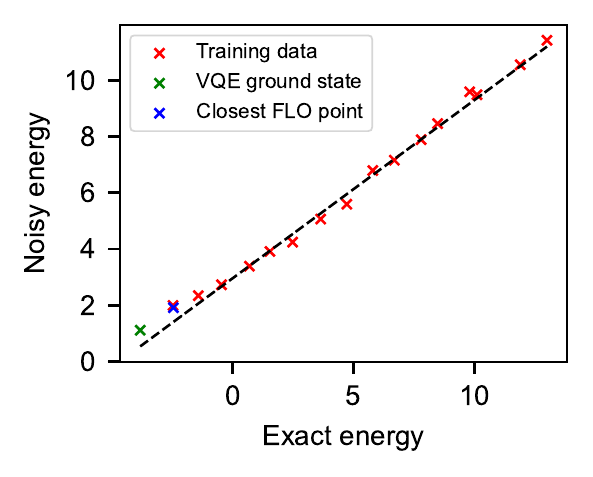}
    \end{minipage}
    \hfill
    \begin{minipage}[c]{0.32\textwidth}
        \subcaption{(c) $1\times4$, $U=4$, depth 2}
        \includegraphics[scale=0.95]{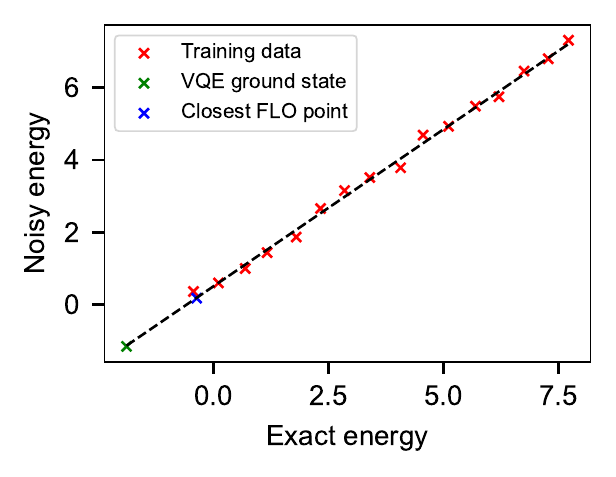}
    \end{minipage}
    \caption{Representative examples showing a near-linear relationship between noisy and exact energies. All results are at half-filling, corresponding to the most complex circuits we execute. The straight line demonstrates the fit found using the Theil-Sen algorithm we used. Residual error is reduced by subtracting the remaining error in the closest FLO point to the VQE parameters (shown in green). Error bars are too small to be visible.}
    \label{fig:tflo}
\end{figure*}

Here we make several optimisations to fine-tune the performance of TFLO for larger system sizes and enable us to achieve high-quality results. As with previous work\cite{montanaro21}, we use an algorithm based on linear regression to infer the map from noisy to real energies. As justification for this, if the intended quantum state produced by the experiment with variational parameters $\mathbf{\theta}$ is $\psi(\mathbf{\theta})$ and the noise process that occurs in the quantum computer is a quantum channel that maps 
\be \label{eq:tflo_decoherence} \psi(\mathbf{\theta}) \mapsto \widetilde{\psi(\mathbf{\theta})} = (1-p)\psi(\mathbf{\theta}) + p \rho \ee
for some fixed quantum state $\rho$, then the noisy energy is a corresponding linear transformation of the exact energy $\tr H \widetilde{\psi(\mathbf{\theta})}$. Some natural error processes occurring in quantum hardware are indeed of this form (such as the depolarising channel applied to the output of the experiment); however, many other errors are not. Nevertheless, in practice we observe an approximately linear relationship between exact and noisy energies (Fig.\ \ref{fig:tflo}).

However, one can see that not all points fit perfectly along a line, and we expect systematic and transient errors to occur that lead to certain data points being low-quality. We use the Theil-Sen estimator\cite{theil50,sen68} to reduce the effect of these outliers, which is a linear regression algorithm based on taking the median of the slopes between pairs of points.

Next, to further improve the tolerance of this method to noise, we look for a fairly large number (here we choose 16) of tuples of parameters whose corresponding energies are well spread-out, to minimise the effect on the linear fit of systematic or transient errors in computing each individual energy. For the case of 1 VQE layer, FLO circuits have two nonzero VQE parameters for $1 \times L_y$, and three nonzero parameters for $2 \times L_y$. We find parameters whose corresponding exact energies are spread out by searching over energies obtained for a uniformly spaced grid of size 16 for each parameter. The cost of this method grows exponentially with the number of layers, so for 2 or more layers, we instead search over 256 random parameter choices. In all cases, we use efficient classical simulation software previously developed for VQE for the Fermi-Hubbard model\cite{cade20} to compute these energies. This code uses an exponential-time simulation approach which does not use the efficient algorithm for simulating FLO circuits\cite{terhal02}; however, for the problem sizes considered, it is sufficiently efficient. As larger problem sizes are considered, it will become essential to use a theoretically efficient classical simulation algorithm.

A final important improvement that we make to the previously developed TFLO algorithm is designed to correct for coherent errors. In principle, TFLO with linear regression can correct for decoherence of the form of (\ref{eq:tflo_decoherence}) with very high accuracy. However, an important class of errors which are not corrected in this way are coherent errors that depend on the choice of parameters (for example, over-rotations). To correct for errors of this form, after performing TFLO as previously discussed, we implement a final step where we subtract off the residual error at the FLO point which is closest to the parameters at which we wish to evaluate the energy -- the point obtained by setting the onsite parameter(s) to zero. The intuition behind this step is that, as almost all gates in this FLO circuit are the same as the real circuit we wish to execute, we expect the error behaviour to be very similar.

\textbf{Other observables.} As well as the energy, TFLO can be applied to any other 
observable, as long as the exact and noisy expectations are available for the FLO points.
Except for the energy, all observables considered by us are diagonal in the 
computational basis, which means their noisy expectation values can be approximately computed from the 
same samples that were used to compute the expectation value of the onsite term in the 
Hamiltonian. For any diagonal observable, the corresponding exact expectation value can be approximately computed from samples in the computational basis at the FLO point, which can be generated efficiently classically\cite{terhal02}; the observables considered here could also be computed exactly. (In our experiments, for ease of implementation we instead used samples generated by simulating the circuit directly, which is sufficiently efficient for the circuit sizes we consider.)

Hence TFLO can be applied to all other observables considered in this paper
as well, without the need for additional quantum resources.
However, a caveat applies: The evaluation points for TFLO were chosen such that 
the  energies are spread out to facilitate a good linear fit. This does not
necessarily imply that the expectation values of other observables are also spread 
out.

In fact, at half-filling the ground state expectation value of all single-site density
$n_{i\sigma}$ operators is uniformly $\frac{1}{2}$ due to the particle-hole
symmetry. And for states with uniform density, the density is invariant under FLO circuits 
(we will prove this shortly).
Hence the inferred linear transformation from noisy densities to exact densities will
always be the constant $\frac{1}{2}$-function. Note however, that in this special case 
of half-filling the particle-hole symmetry holds regardless of whether $U=0$ and hence
predicting $\braket{n_{i\sigma}} = \frac{1}{2}$ is actually correct. For other observables (e.g. charge densities) 
the expectation value of the observable to mitigate is invariant under FLO circuits, but does change when applying non-FLO gates. In such cases TFLO would fail altogether and would always predict the value of the FLO simulations. To avoid these issues, we performed the following checks. First, if the exact observable values found by classical simulation were all close (within 0.05), we only applied the coherent error correction step, and not a linear fit (which would be meaningless in this case). Otherwise, we performed a linear fit using the Theil-Sen algorithm, and checked whether the linear fit was a good explanation of the data, as measured by the coefficient of determination ($R^2$) being larger than 0.7. If this check failed, we assume that there is no simple relationship between the noisy and exact values and simply return the noisy value of the observable.

The fact that the charge and spin densities of states with constant density are invariant under FLO 
circuits can be shown as follows. First note that any (number-preserving) FLO unitary can be written as the 
product of Givens rotations, hence it suffices to show the statement for Givens rotations.
Without loss of generality, consider a Givens rotation applied to the first two qubits, write the state as 
\begin{equation}
    \ket{\Psi} = \ket{00} \ket{\psi_{00}} + \ket{01} \ket{\psi_{01}}
               + \ket{10} \ket{\psi_{10}} + \ket{11} \ket{\psi_{11}}
\end{equation}
and note that $\braket{n_1}_\Psi = \braket{n_2}_\Psi$ implies
$\braket{\psi_{01}|\psi_{01}} + \braket{\psi_{11}|\psi_{11}} = \braket{\psi_{10}|\psi_{10}} + \braket{\psi_{11}|\psi_{11}} \Leftrightarrow \braket{\psi_{01}|\psi_{01}} = \braket{\psi_{10}|\psi_{10}}$.
Applying a Givens rotation with angle $\theta$ to the first two qubits then yields
\begin{equation}
\begin{aligned}
    \ket{\Psi'} &= G_{12}(\theta) \ket{\Psi} \\
                &= \ket{00} \ket{\psi_{00}}
                + [\cos(\theta) \ket{01} + \sin(\theta) \ket{10}] \ket{\psi_{01}} \\
                &+ [-\sin(\theta) \ket{01} + \cos(\theta) \ket{10}] \ket{\psi_{10}}
                + \ket{11} \ket{\psi_{11}}
\end{aligned}
\end{equation}
and computing  the density on the first qubit then yields
\begin{equation}
\begin{aligned}
    \braket{n_1}_{\Psi'} &= \cos(\theta)^2 \braket{\psi_{01}|\psi_{01}} 
                       + \sin(\theta)^2 \braket{\psi_{10}|\psi_{10}} \\
                       &+ \braket{\psi_{11}|\psi_{11}}
                       = \braket{n_1}_\Psi
\end{aligned}
\end{equation}
where we used for the second equality that
$\braket{\psi_{01}|\psi_{01}} = \braket{\psi_{10}|\psi_{10}}$.
Similarly $\braket{n_2}_{\Psi'} = \braket{n_2}_\Psi$. For all other sites $j$ the density is invariant because $[n_j, G_{12}(\theta)] = 0$.

%-------------------------------------------

\begin{figure*}
    \centering
    \begin{minipage}[c]{0.32\textwidth}
        \subcaption{(a) Chemical potential, $U=4$}
        \includegraphics[scale=.95]{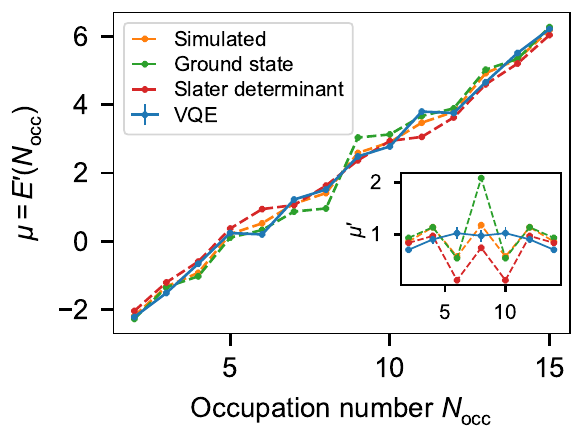}
    \end{minipage}
    \begin{minipage}[c]{0.32\textwidth}
        \subcaption{(b) Chemical potential, $U=8$}
        \includegraphics[scale=.95]{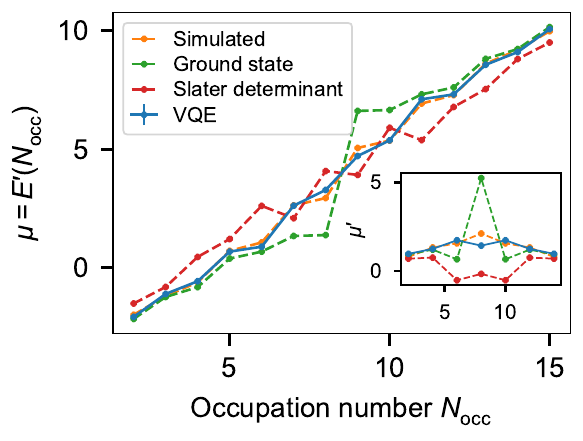}
    \end{minipage}
    \begin{minipage}[c]{0.32\textwidth}
        \subcaption{(c) Staggered spin correlation}
        \includegraphics[scale=.95]{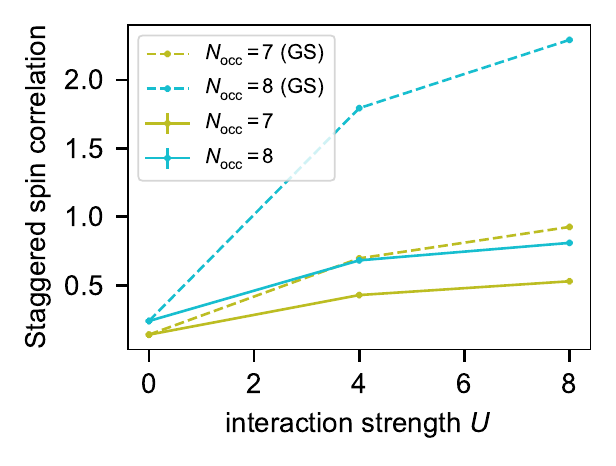}
    \end{minipage}
    \caption{Additional results for $2\times 4$. (a)--(b): Chemical potentials $\mu(\Nocc) = E(\Nocc)-E(\Nocc-1)$. Labels are as in Fig.\ \ref{fig:energies}. Inset shows the derivative $\mu'(\Nocc) = E(\Nocc+1)-2E(\Nocc)+E(\Nocc-1)$ of the chemical potential at even occupations. (c): Total staggered spin correlation $\sum_s (-1)^s \braket{S_s^z S_{s+1}^z}$ at and close to half-filling, where the ordering is taken along the Jordan-Wigner snake (Fig.\ \ref{fig:2x4_swap}(a)). It compares experimental results from VQE with true values in the ground state. $U=0$ points are exact classical calculations.}
    \label{fig:mott2x4}
\end{figure*}

\section{Classical comparators}
\label{app:data_analysis}

In this appendix we give more details about how the classical comparator quantities in Figures \ref{fig:energies}, \ref{fig:densities_spins}, \ref{fig:spin_correlations} and elsewhere were calculated.

\textbf{Simulated and ground state quantities.} The ``simulated'' line in \ref{fig:energies} refers to the energy produced using an exact classical simulation of the VQE algorithm, using a software package targeted at efficient simulation of the Fermi-Hubbard model\cite{cade20}. The optimisation algorithm used was BFGS, which was previously found to be effective\cite{cade20}, but can get trapped in local minima; results were therefore checked using the Cobyla algorithm. Ground state quantities such as energies, charges, spins, and correlations were computed by exact diagonalisation.

\textbf{Slater determinant.} A natural classical comparator against the performance of the variational ansatz of quantum states that we use is the family of Slater determinant (Hartree-Fock) states, which can be seen as product states that obey appropriate fermionic antisymmetry.

Given a Slater determinant, its energy with respect to the Fermi-Hubbard Hamiltonian can be computed efficiently either using a general technique for simulating fermionic linear optics circuits\cite{terhal02}, or more simply via the Slater-Condon rules from quantum chemistry. One particular Slater determinant that  can be used as a trial state is the ground state of the $U=0$ (noninteracting) Fermi-Hubbard model. However, it is possible to achieve an energy closer to the true ground energy, for example by the well-known mean-field approximation to the Fermi-Hubbard model\cite{kadanoff09}.

The iterative mean-field approach is not guaranteed to converge, or to find a Slater determinant that minimises the energy. To measure the ability of a ``best possible'' Slater determinant  to compete with the VQE solution, we therefore used a different approach, where we optimised (classically) over the space of Slater determinants, with the goal of minimising the energy with respect to the Fermi-Hubbard Hamiltonian. To parametrise this space, we used the entries of an $L\times L$ Hermitian matrix $h$ for a system with $L$ sites, corresponding to a Hamiltonian $H_{SD} = \sum_{i,j} h_{ij} a_i^\dag a_j$, with the same matrix $h$ being used for spin-up and spin-down, to ensure that spin-flip symmetry was obeyed. Then the Slater determinant with occupation number $k$ corresponding to this matrix, the ground state of $H_{SD}$, is found by taking the $k$ eigenvectors $e_i = (\alpha_{i1},\dots,\alpha_{iL})$ of $h$ with lowest eigenvalues and forming the product of single-particle operators $\alpha_{i1} a^\dag_1 + \dots + \alpha_{iL} a^\dag_L$. In the case of even occupation numbers, we used the same occupation number for spin-up and spin-down; for odd occupation numbers, we had one more spin-up electron (matching the VQE experiments).

We used the Slater-Condon rules to compute the overall Fermi-Hubbard energy corresponding to $h$, as a subroutine within the classical BFGS optimiser. This allowed us to minimise this energy over $h$, for several randomly perturbed starting conditions. We found that this approach reliably converged to good solutions.

%-------------------------------------------

\section{Error analysis}
\label{app:error_analysis}

Error bars for energies and other quantities computed using VQE were derived as follows. First, we assume that measurements of each observable -- conditioned on the occupation number in each spin sector being correct -- can be modelled by a Gaussian distribution. We approximate the mean and variance of this distribution by the sample mean and sample variance found experimentally. We then need to take into account additional variance coming from the uncertainty in the number of runs retained after postselection. With $N$ trials in total, standard deviation $\sigma$ (after postselecting), and probability $p$ of postselection, it turns out\cite{marciniak99} that the variance of the sample mean is $(\sigma^2 / (pN))(1 + (1 - p)/(pN) + O(1/(pN)^2))$ (see Section \ref{sec:variance_postselection} for a proof).

We now have error bars for the ``raw'' observable values produced after postselection, but before the other error mitigation techniques. As it is not straightforward to understand the effect of the TFLO procedure on errors analytically, we produce error bars for observables after TFLO using a Monte Carlo technique, where we assume that raw observables are distributed according to Gaussians with means and variances determined by the previous step. We then sample observables from these distributions 1000 times for each of the parameter settings used in TFLO (i.e.\ the FLO points and the VQE ground state point) and run the TFLO procedure to produce an energy estimate. The error bar we report is then the sample standard deviation of this estimate.

In the cases of quantities derived from expectations of multiple observables (i.e.\ spin and charge correlations) we make the simplifying assumption that the distribution of each of the observables combined to produce that quantity is independent to produce an overall error bar.

Error bars for the energies reported by BayesMGD are the internal estimates produced as described in Appendix \ref{sec:bayesmgd-appendix}. These show the level of certainty of the algorithm but may not correspond to a true error bar for the energy, if it were measured at the current parameters.

\subsection{Variance of observables taking postselection into account}
\label{sec:variance_postselection}
First, we calculate an expression stated in Ref.~\cite{marciniak99} for the variance of the sample mean when the number of samples is random.
Assume that $X_1, \cdots, X_N$ are $N$ i.i.d random variables with mean $\mu$ and variance $\sigma^2$ and $Y_1, \cdots, Y_N$ are $N$ i.i.d. Bernoulli variables with $p(Y_i = 1) = p$ for all $i$. The $X_i$ play the role of our samples and $Y_i$ indicates if we keep the $i$-th sample after postselection or not. The estimator for the mean $\mu$ is then
\begin{equation}
    Z = \frac{\sum_i Y_i X_i}{\sum_j Y_j}.    
\end{equation}
It is straightforward to verify that this estimator is unbiased:
\begin{equation}
    \mathds{E}[Z] = \mathds{E}\left[ \frac{\sum_i Y_i X_i}{\sum_i Y_i}\right] 
                  = \sum_i \mathds{E}[X_i] \mathds{E}\left[ \frac{Y_i}{\sum_j Y_j}\right] 
                  = \mu.
\end{equation}
To calculate its variance, it is useful to condition on the number of successful samples
$|Y| = \sum_i Y_i$:
%Letting $|Y|$ denote $\sum_i Y_i$, we have
\begin{equation}
\begin{aligned}
\mathds{E}[Z^2] &= \sum_y \frac{\Pr[Y=y]}{|y|^2} \mathds{E}_X[(\sum_{i,y_i=1} X_i)^2]\\
&= \sum_y \frac{\Pr[Y=y]}{|y|^2} \mathds{E}_X[\sum_{i,y_i=1} X_i^2 + \sum_{i\neq j, y_i=y_j=1} X_iX_j ]\\
&= \sum_y \frac{\Pr[Y=y]}{|y|^2} |y| \mathds{E}[X_0^2] + |y|(|y|-1) \mathds{E}[X_0]^2 \\
&= \mathds{E}[X_0]^2 + \sum_y \frac{\Pr[Y=y]}{|y|} ( \mathds{E}[X_0^2] - \mathds{E}[X_0]^2),
\end{aligned}
\end{equation}
so
\begin{equation}
    \operatorname{Var}(Z) = \mathds{E}[Z^2] - \mathds{E}[Z]^2 = \sigma^2 \mathds{E}_Y[1/|Y|].
\end{equation}
To calculate the remaining expectation value, note that in all our cases $|Y|$ is large
(around 20,000) and hence well concentrated. A straightforward Taylor expansion of 
$f(|Y|) = \frac{1}{|Y|}$ around its expectation value $Np$ yields then
\begin{equation}
    \frac{1}{|Y|} = \frac{1}{Np} - \frac{|Y| - Np}{(Np)^2} + \frac{(|Y| - Np)^2}{(Np)^3}
    + O\left((Np)^{-4}\right)
\end{equation}
and taking expectation values on both sides gives
\begin{equation}
    \mathds{E}_Y \left[\frac{1}{|Y|}\right] = \frac{1}{Np} - 0 + \frac{N p (1-p)}{(Np)^3}
\end{equation}
which finally makes the variance
\begin{equation}
    \operatorname{Var}(Z) = \frac{\sigma^2}{Np} \left(1 + \frac{1-p}{Np} \right) 
    + O\left((Np)^{-3}\right).
\end{equation}
Several more complicated and accurate asymptotic expansions are given in Ref.~\cite{marciniak99} and references therein. The above expression is sufficient for our needs (indeed, the $O(1/(pN)^2)$ correction is already almost unnoticeable).

%-------------------------------------------

\section{Additional results for the $2\times 4$ lattice}
\label{app:2x4_figures}

\begin{figure*}
    %\centering
    \hspace{-2cm}
    \begin{minipage}[c]{0.35\linewidth}
    \subcaption{(a) VQE, $U=4$}
    \includegraphics[height=3cm]{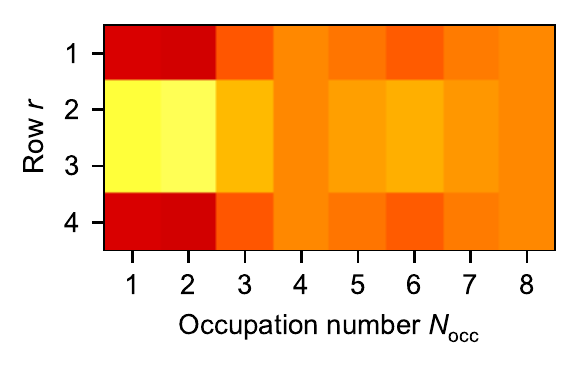}\\
    \subcaption{(e) Ground state, $U=4$}
    \includegraphics[height=3cm]{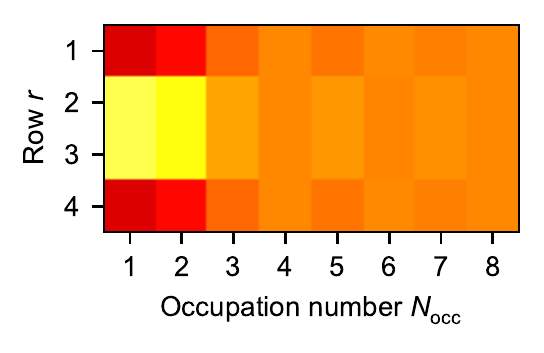}
    \end{minipage}
    \hspace{-0.8cm}
    \raisebox{0.5cm}{\begin{minipage}[l]{0.02\linewidth}
    \includegraphics[height=5.5cm]{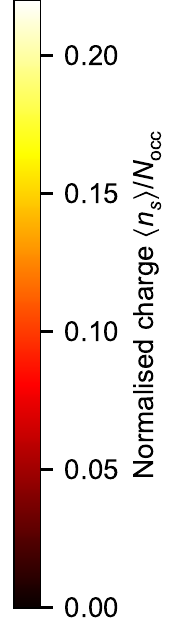}
    \end{minipage}}
    \hspace{2cm}
    \begin{minipage}[l]{0.15\linewidth}
    \subcaption{(b) VQE, $U=4$, $\Nocc$ even}
    \includegraphics[height=3cm]{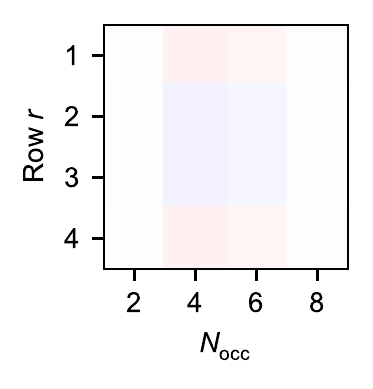}\\
    \subcaption{(f) Ground state, $U=0$, $\Nocc$ odd}
    \includegraphics[height=3cm]{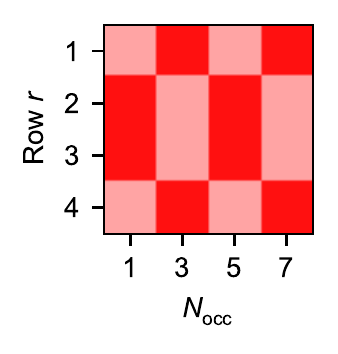}
    \end{minipage}
    \hspace{0.2cm}
    \begin{minipage}[l]{0.15\linewidth}
    \subcaption{(c) VQE, $U=4$, $\Nocc$ odd}
    \includegraphics[height=3cm]{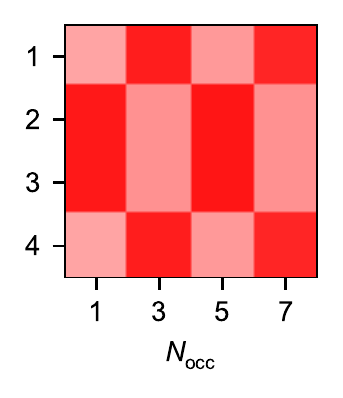}\\
    \subcaption{(g) Ground state, $U=4$, $\Nocc$ odd}
    \includegraphics[height=3cm]{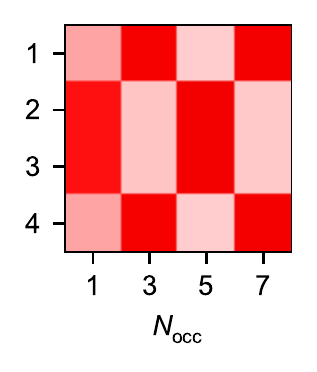}
    \end{minipage}
    \begin{minipage}[l]{0.15\linewidth}
    \subcaption{(d) VQE, $U=8$, $\Nocc$ odd}
    \includegraphics[height=3cm]{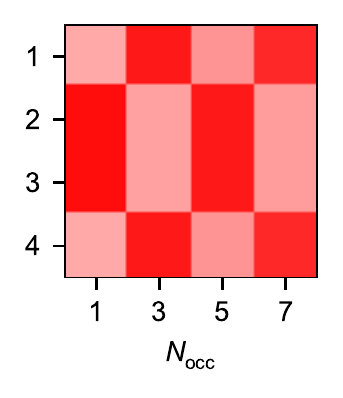}\\
    \subcaption{(h) Ground state, $U=8$, $\Nocc$ odd}
    \includegraphics[height=3cm]{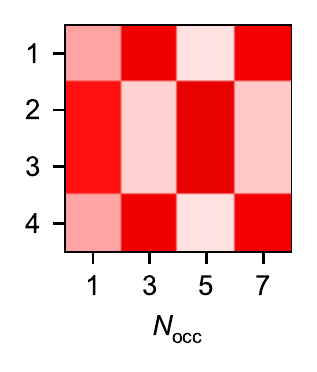}
    \end{minipage}
    \raisebox{0.8cm}{\begin{minipage}[l]{0.02\linewidth}
    \includegraphics[height=5.5cm]{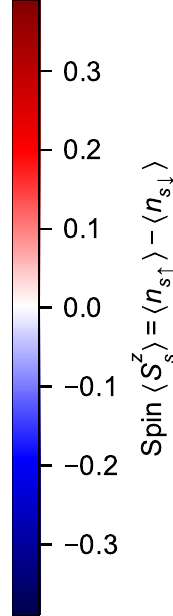}
    \end{minipage}}
    \caption{Charge and spin densities for a $2\times 4$ lattice. Top row: Experimental VQE results. Bottom row: Ground state. X axis: occupation number $\Nocc$; Y axis: row index. By symmetry, charges and spins are equal for each pair of sites in each row, so plot shows charge/spin for one site in each row. Spin plots split by even/odd occupations. In the ground state, spin is 0 everywhere for $\Nocc$ even.}
    \label{fig:2x4_densities_spins}
\end{figure*}

In this appendix we collate figures and additional discussion for our results on a $2\times 4$ lattice.

\textbf{Metal-insulator transition.} In Fig. \ref{fig:mott2x4} (a)-(b), we observe the behaviour of the chemical potential for different occupations. We do not see the onset of the MIT, as $\mu'$ remains essentially constant. We suspect that this is due to two compounded effects: uncorrected errors affecting the quantum processor, and the low depth of the EHV ansatz used not being able to sufficiently capture correlations in this system, as we discuss below.

\textbf{Charge density expectation.} In large 2D systems, it is expected that at half-filling the charge becomes commensurate with the lattice, spontaneously breaking discrete translation symmetry. With this in mind we study the charge density in the $2\times 4$ system (Fig. \ref{fig:2x4_densities_spins}(a),(e)). In this small system we find that the charge expectation inherits some of the physics of 1D. A way of seeing this is the following. Consider the non-interacting limit $U=0$ with zero hopping along the rungs $(t_R=0)$. In this case the energy of the system consists of two degenerate bands each corresponding to a $1\times 4$ system. Increasing the hopping along the rungs to its original $t_R=1$ value, the two degenerate bands split in energy by $t_R$. The single particle states in the lower band correspond to the symmetric combinations of electrons from both chains. Increasing the interactions from zero gradually, the system resembles a $1\times 4$ Hubbard chain of fermions delocalised along the rungs. This explains the uniform charge density profile at a quarter filling ($\Nocc=4$), as this corresponds to a completely filled lower band of ``rung'' electrons. This delocalisation also decreases the effect of the onsite interaction, as can be seen  by comparing a true 1D system with the effective one made of ``rung'' electrons using standard perturbation theory around the non-interacting regime.

\textbf{Spin density.} The spin profiles for odd occupations for the VQE ansatz (Fig.~ \ref{fig:2x4_densities_spins}(c)--(d)) and for the true ground state (Fig.~\ref{fig:2x4_densities_spins}(g)-(h)) are similar, and close to the non-interacting pattern (Fig.~\ref{fig:2x4_densities_spins}(f)). This supports the interpretation that VQE is able to capture non-correlated behaviour at depth 1 in this system, to reproduce the physics of a weakly interacting system, but fails to account for correlations, which in this system are ``screened'' by the delocalisation of particles across the rungs. Based on this, we expect that to observe truly correlated behaviour in this system (e.g.\ a peak in $\mu'$ around half-filling, Fig.~\ref{fig:2x4_densities_spins}), we need a deeper VQE circuit to represent the ground state observables with higher accuracy.

%-------------------------------------------

\section{Further details about BayesMGD}
\label{sec:bayesmgd-appendix}

In Appendix \ref{sec:variational-optimiser}, in particular Eq.~\eqref{eq:bayesmgd-update}, an explicit
formula for $\vec\beta_m$ and $\vec\Sigma_m$ is still missing. We give these explicit 
formulas here together with their derivation and also present pseudo-code for the
BayesMGD algorithm and show how it is related to Kalman filters.

\subsection{The Bayesian update step}
\label{sec:bayesmgd_update_step}
We make three main assumptions in our design of the BayesMGD algorithm:
\begin{enumerate}
    \item For all $\vec\theta'$ near $\vec\theta_{m-1}$ the probability of observing
    some value $y'$ is given by
    $p(y'\,|\,\vec\theta', \vec\beta) = \mathcal{N}(f_s(\vec\theta'; \vec\beta))$ where 
    the surrogate model can be written as
    $f_s(\vec\theta'; \vec\beta) = \sum_j^{n_m} \beta_j \phi_j(\vec\theta)$ with $n_m$ \emph{model functions} $\phi_j(\vec\theta)$ that can be read off Eq.~\ref{eq:surrogate_model}.
    \item In the $m$-th iteration, before observing new data our belief about the model 
    parameters is a multivariate Gaussian
    $p_{m|m-1}(\vec\beta) = \mathcal{N}(\vec\beta_{m|m-1}, \vec\Sigma_{m|m-1})$.
    \item When making a gradient descent step, we lose certainty about the 
    model parameters $\vec\beta$ proportional to the step width $s$, but our belief
    does not change. That is, making the gradient descent step sends
    $\vec\Sigma_{m+1|m} = \vec\Sigma_{m|m} + \frac{s^2}{l^2} \mathds{1}$ and
    $\vec{\beta}_{m+1|m} = \vec{\beta}_{m|m} $.
\end{enumerate}
Assumptions 1 and 2 imply that after observing new data $\{y_i,\sigma_i\}$ at points
$\{\vec\theta^{(i)}\}$ we can use Bayes' theorem (compare \eqref{eq:bayesmgd-update})
to compute the posterior of $\vec\beta$,
\begin{equation}
\begin{aligned}
  p_{m|m}(\vec\beta) \propto& P(\{y_i, \sigma_i\}\,|\, \{\vec\theta^{(i)}\}, \vec\beta)
                  \, p_{m|m-1}(\vec\beta) \\ 
                 =& \prod_{i=1}^p \mathcal{N}(y_i; f_s(\vec\theta^{(i)}; \vec\beta), \sigma_i)  \\
                    & \times 
                    \mathcal{N}(\vec\beta ; \vec\beta_{m|m-1}, \vec\Sigma_{m|m-1}) \\
                 =& \mathcal{N}(\vec\beta ; \vec\beta_{m|m}, \vec\Sigma_{m|m}),
\end{aligned}
\end{equation}
Defining the \emph{design matrix}
\begin{equation}
    \vec X = \begin{pmatrix}
        \phi_1(\vec\theta^{(1)}) & \cdots & \phi_n(\vec\theta^{(1)}) \\ 
        \vdots &   & \vdots \\ 
        \phi_1(\vec\theta^{(p)}) & \cdots & \phi_n(\vec\theta^{(p)}) \\ 
    \end{pmatrix},
\end{equation}
the \emph{measurement noise matrix}
\begin{equation}
    \vec\Sigma = \begin{pmatrix}
               \sigma_1^2 & & \\
                & \ddots & \\
                & & \sigma_p^2 \\ 
             \end{pmatrix}
\end{equation}
and a \emph{measurement outcome vector} 
\begin{equation}
    \vec y = (y_1, \cdots, y_p)^\dagger,
\end{equation}
and plugging these into the expressions for Gaussian distibutions yields
\begin{equation}
\begin{aligned}
    \log p_{m|m}(\vec\beta) =&
        - \frac{1}{2} (\vec y - \vec X \vec\beta)^\dagger \vec\Sigma^{-1} (\vec y - \vec X \vec\beta) \\ 
        &-\frac{1}{2} (\vec\beta - \vec\beta_{m|m-1})^\dagger \vec\Sigma_{m|m-1}^{-1} (\vec\beta - \vec\beta_{m|m-1}) \\
        &+ \mathrm{const},
\end{aligned}
\end{equation}
where the normalisation factors were absorbed into ``const''.
The product of two Gaussian distributions is again a Gaussian distribution. So 
we know that 
\begin{equation}
    \log p_{m|m}(\vec\beta) =
    -\frac{1}{2} (\vec\beta - \vec\beta_{m|m})^\dagger \vec\Sigma_{m|m}^{-1} (\vec\beta - \vec\beta_{m|m})
                  + \mathrm{const}
\end{equation}
for some $\vec\beta_{m|m}$ and $\vec\Sigma_{m|m}$. Comparing terms in the last two
equations then gives
\begin{equation}
\begin{aligned}
    \vec\Sigma_{m|m}^{-1} &= \vec X^\dagger \vec\Sigma^{-1} \vec X
                       + \vec\Sigma_{m|m-1}^{-1} \\ 
    \vec\beta_{m|m} &= \vec\Sigma_{m|m} (\vec X^\dagger \vec\Sigma^{-1} \vec y + \vec\Sigma_{m|m-1}^{-1} \vec\beta_{m|m-1}).
\end{aligned}
\label{eq:bayesmgd_update_ii}
\end{equation}

\subsection{Pseudocode for BayesMGD}

In Algorithns \ref{alg:bayesmgd} and \ref{alg:bayesupdate} we give pseudocode for the 
BayesMGD algorithm and the Bayesian update step. Throughout we assume a linear
model of the form 
$
  f_s(\vec\theta; \vec\beta) = \sum_{j=1}^{n_m} \beta_j \phi_j(\vec\theta),
$
where $n_m$ is the number of model parameters and the model is linear in the model
parameters $\beta_j$, but the model functions $\phi_j$ are not necessarily linear in $\vec\theta$. Compare
this with eq.~\eqref{eq:surrogate_model} to identify the model functions
$\phi_j(\vec\theta)$. For simplicity of notation, we also implicitly lift any function 
that is passed a variable together with the uncertainty of the variable to a version
that returns the function value together with its uncertainty computed using Gaussian
error propagation.

\SetInd{0.5em}{0.7em}
\begin{algorithm}[t]
\SetAlgoLined
\DontPrintSemicolon
  \KwIn{function $f$ returning uncertain values $y, \sigma_y$,\\
           model functions $\{\phi_j\}_{j=1}^{n_m}$,\\
           initial point $\vec\theta$, stability constant $A$,\\
           learning rate $\gamma$, rate decay exponent $\alpha$,\\
           sample number $p$, maximum evaluations $n_{\mathrm{eval}}$,\\
           tolerance $\epsilon$, length scale $l$, \\
           initial parameters $\vec\beta$, covariance matrix $\vec\Sigma$
        }
        
  $m \leftarrow 0$ \;
  \While{$m < n_{\mathrm{eval}}$}{
    $m \leftarrow m+1$ \;
    $\delta_m \leftarrow \delta / m^\xi$  \Comment*[r]{Set sample radius}
    $\gamma_m \leftarrow \gamma / (m+A)^\alpha$    \Comment*[r]{and step width}
    $S \leftarrow $ sample $p$ points from $\delta_m$-ball around $\vec\theta$ \;
    $L \leftarrow \{\}$\;
    \For{$\vec\theta_i \in S$}{
      Add $f(\vec\theta_i) = (y_i, \sigma_i)$ to $L$\;
    }
    $(\vec\beta, \vec\Sigma) \leftarrow \textrm{BayesUpdate}(\{\phi_j\}, S, L, \vec\beta, \vec\Sigma)$\;
    $\vec g \leftarrow \nabla_{\vec\theta}f_s(\vec\theta ; \vec\beta)$ \;
    $\vec\theta \leftarrow \vec\theta - \gamma_m \vec g$     \Comment*[r]{Gradient descent step}
    $\vec\Sigma \leftarrow \vec\Sigma + \frac{|\gamma_m \vec g|^2}{l^2}\mathds{1}$
    \Comment*[r]{Add uncertainty due to step}
    \If{$\gamma_m \cdot |\vec g| < \epsilon$}{ 
      $(y, \sigma_y) \leftarrow f_s(\vec \theta\vec;\; \vec\beta, \vec\Sigma)$\;
      \Return{$((y, \sigma_y), \vec\theta)$} \Comment*[r]{Return if step size is small}
    }
  }
  $(y, \sigma_y) \leftarrow f_s(\vec \theta\vec; \vec\beta, \vec\Sigma)$\;
  \Return{$((y, \sigma_y), \vec\theta)$} \Comment*[r]{Return if maximum iterations reached}
\caption{BayesMGD}
\label{alg:bayesmgd}
\end{algorithm}

\SetInd{0.5em}{0.7em}
\begin{algorithm}[t]
\SetAlgoLined
\DontPrintSemicolon
  \KwIn{model functions $\{\phi_j\}_{j=1}^{n_m}$,\\
        evaluation points $\{\vec\theta_i\}_{i=1}^N$,\\
        values with noise $\{(y_i, \sigma_i)\}_{i=1}^N$,\\
        prior parameters with covariance $\vec\beta_0, \vec\Sigma_0$
        }
    Create empty $n_m \times N$ matrix $\vec X$ 
    \Comment*[r]{Data preparation}
    $X_{ji} \leftarrow \phi_j(\vec\theta_i)$ \;
    Collect the $y_i$ into $\vec y$\;
    Collect the $\sigma_i^2$ onto diagonal of $\vec\Sigma$ \;
    $\vec\Sigma_1^{-1} \leftarrow \vec X^\dagger \vec\Sigma^{-1} \vec X + \vec\Sigma_0^{-1}$ 
    \Comment*[r]{Update equations}
    $\vec\beta_1 \leftarrow \vec\Sigma_1 (\vec X^\dagger \vec\Sigma \vec y
                                           + \vec\Sigma_0^{-1}\vec\beta_0)$\;
    \Return{$(\vec\beta_1, \vec\Sigma_1)$}
\caption{BayesUpdate}
\label{alg:bayesupdate}
\end{algorithm}

\subsection{Relation between BayesMGD and Kalman filters}
\label{sec:bayesmgd-kalman-filters}

The Kalman filter is an algorithm that iteratively combines noisy measurement data
with prior knowledge and knowledge of the dynamics to estimate the state of a 
dynamical system\cite{kalman60}. A notable, early application
was in the Apollo Guidance Computer \cite{apolloguidance69}. 
However, as far as we know, Kalman filters have not yet been considered in the context of function 
optimisation. Instead their main application so far was to estimate the state of physical systems, like
air- or spacecraft and robotics.
We will now show that our BayesMGD algorithm is mathematically equivalent to the Kalman filter algorithm.

In the language of Kalman filters,
\begin{itemize}
    \item $\vec x_{m-1|m-1}$ is the state estimate at time step $m$ given all
    measurements up to and including time step $m-1$. In our case these are the 
    most likely model parameters $\vec\beta_{m-1|m-1}$;
    \item $\vec P_{m-1|m-1}$ is the covariance matrix of that estimate. In our case 
    this is $\vec\Sigma_{m-1|m-1}$.
\end{itemize}
From these the state at time $m$ is predicted as 
\begin{equation}
\begin{aligned}
    \vec x_{m|m-1} &= \vec F_m \vec x_{m-1|m-1} + \vec B_m \vec u_m \\
    \vec P_{m|m-1} &= \vec F_m \vec P_{m-1|m-1} \vec F_m^\dagger + \vec Q_m
\end{aligned}
\end{equation}
where $\vec{F}_m$ encodes the dynamics of the system, $\vec B_m \vec u_m$ is the 
result of external control inputs and $\vec Q_m$ is the process noise. In our 
case the dynamics are trivial ($\vec F_m = \mathds{1}$), there is no external 
control input $\vec u_m$ and the process noise is $\vec Q_m = \frac{s^2}{l^2}$.
In the next step, observation data is used to refine the prediction. The 
observations $\vec z_m$ are linearly related to the true state $\vec x_m$ as 
\begin{equation}
    \vec z_m = \vec H_m \vec x_m + \vec v_m
\end{equation}
where $\vec{H}_m$ is some (possibly $m$-dependent) matrix and $\vec v_m$ the observation noise which
is assumed to be zero-mean gaussian white noise $\vec v_m \sim \mathcal{N}(\vec 0, \vec R_m)$.
In our case the observation data are the 
measurement outcomes $\vec y$, the measurement noise is $\vec \Sigma$ and the
design matrix $\vec X$ plays the role of $\vec H_m$. Using this data the
prediction is refined to 
\begin{equation}
\begin{aligned}
    \vec P_{m|m}^{-1} &= \vec H^\dagger_m \vec R_m^{-1} \vec H_m
                       + \vec P_{m|m-1}^{-1} \\
    \vec x_{m|m} &= \vec P_{m|m} (\vec H^\dagger_m \vec R_m^{-1} \vec z_k
                                 + \vec P_{m|m-1}^{-1} \vec x_{m|m-1}).
\end{aligned}
\end{equation}
With the above identifications these are the same equations as
\eqref{eq:bayesmgd_update_ii} and it is clear that the way BayesMGD predicts the 
optimal model parameters $\vec\beta$ is the same way that a Kalman filter predicts the system 
state $\vec x$.
 
\subsection{Experimental comparison of optimisation algorithms}
\label{sec:experimental-comparison-of-optimisation-algorithms}

To compare the performance of BayesMGD, vanilla MGD and SPSA we ran VQE for different, well
understood problem instances on real hardware with all three 
optimisation algorithms.

For BayesMGD, we used hyperparameters $A=1$, $p = 1.5 \, \mathrm{dim}(\vec\beta)$, $\gamma=0.3$,
$\alpha=0.602$, $\xi=0.101$, $\delta=0.6$ and $l = 0.2$, which gave consistently good 
results in simulations and were used in the experiments. We implemented MGD based on the description in Ref.~[\onlinecite{sung20}]. MGD has the same hyperparameters as BayesMGD, except for $l$, which is not a hyperparameter. In our experiments we used the same hyperparameter settings as for BayesMGD.
$\vec\beta$ and $\vec\Sigma$ were initialised as %
\begin{equation}
\begin{aligned}
    \vec\beta_0 &= \vec0 \\
    \vec\Sigma_0 &= \mathrm{diag}(
    \underbrace{10^7}_{\Sigma_{\beta_0}},
    \underbrace{10^7, \cdots, 10^7}_{\Sigma_{\beta_j}},
    \underbrace{10^5, \cdots, 10^5}_{\Sigma_{\beta_{jk}}}
    ),
\end{aligned}
\end{equation}
which has the effect that if the least squares fit is underdetermined the algorithm
will first fit with a linear model and only set the $\beta_{jk}$ to non-zero
values if the data cannot be explained with a linear model.

Our implementation of SPSA was the same as that presented in Ref.~\onlinecite{cade20}, except that we only implemented a simple one-stage algorithm, not the three-stage approach used there, as we found that this was sufficient to obtain good performance. We used the same hyperparameters as that previous work, except that we reduced the stability constant $A$, as this was found to be more effective in experiments (we set $\alpha=0.602$, $\gamma=0.101$, $a=0.2$, $c=0.15$, $A=1$).

As a figure of merit
we show the energy difference between the exact energy expectation value one would obtain on a 
perfect device and the best exact energy attainable with the given VQE circuit. For a fair comparison, the number of 
shots per evaluation and number of evaluations per iteration was chosen such that the number of shots
per iteration was the same for all optimisation runs shown in Fig.\ \ref{fig:optimiser_comparison}.

\begin{figure}
    \centering
    \includegraphics{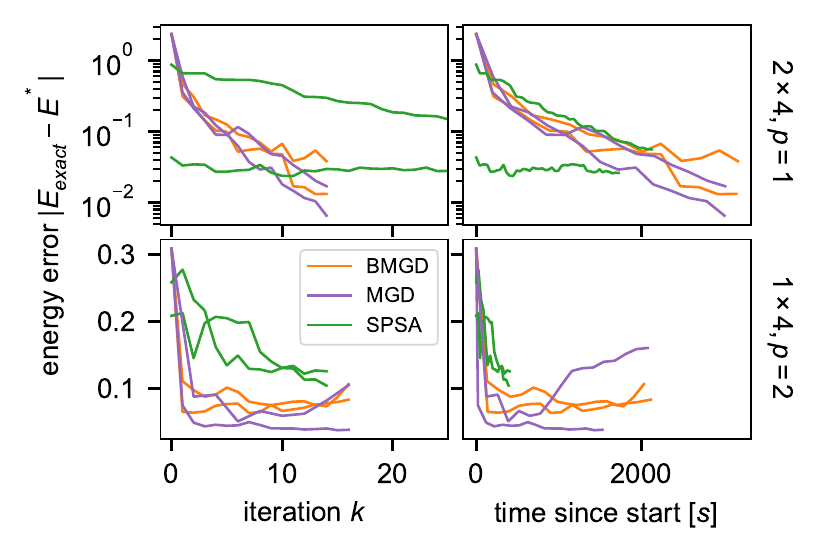}
    \caption{Comparison of the B(ayes)MGD, MGD and SPSA optimisation algorithms. Shown is the energy difference between the classically evaluated, exact energies $E_\mathrm{exact}$ at the parameters produced at each step of the respective optimisation algorithm from noisy measurements on real hardware, and the best attainable energy at the given circuit depth $E^*$, as a function of the optimiser iteration (left column) and 
    as a function of wall clock time since the start of the VQE run (right column). The upper row shows the results on a $2\times 4$ lattice with ansatz depth 1 and the lower row shows the results on a $1\times 4$ lattice with ansatz depth 2. Two runs are shown for each optimisation algorithm and lattice.
    }
    \label{fig:optimiser_comparison}
\end{figure}

The left column of Fig.\ \ref{fig:optimiser_comparison} shows that when the optimisation budget is given in
shots instead of time, BayesMGD and MGD clearly outperform SPSA. However, when the optimisation
budget is wall clock time, the rate of convergence of SPSA and (Bayes)MGD was roughly the 
same, as is best seen in the upper right panel of Fig.\ \ref{fig:optimiser_comparison}. 
In one of the SPSA runs on the $2 \times 4$ lattice the different sources of randomness in the 
SPSA algorithm conspired to send the parameters very close to the global minimum already in the 
first iteration. These are the lower, orange, almost constant lines in the upper row and
interestingly SPSA does not improve the parameters past that, while (Bayes)MGD was able to find 
better parameters even closer to the minimum. The lower row shows that, for harder instances with
more free parameters (here 6 instead of 4), also (Bayes)MGD may fail to improve the parameters
further once it got close to the minimum.

In the instances shown here, there was no clear difference between the performance of BayesMGD and MGD.
However, in simulated tests of VQE for the antiferromagnetic Heisenberg model on the kagome lattice
\cite{bosse21} with 12 qubits and 18 parameters we found that for
$\eta := \tfrac{p}{(n_c+1)(n_c+2)/2} \geq 1$ the performance
of BayesMGD and MGD is comparable, while for $\eta < 1$ BayesMGD often outperforms MGD. As in Appendix \ref{sec:variational-optimiser}, $\eta$ is defined as the ratio between the number $p$ of evaluation points taken in each iteration
and the number $(n_c+1)(n_c+2) / 2$ of evaluation points necessary for a fully determined quadratic fit.
Hence $\eta < 1$ corresponds to an underdetermined quadratic fit, where good usage of prior 
information is especially important.
These situations are shown in Fig. \ref{fig:optimiser_comparison_ii}, where for $\eta = 0.7$ and 
$1.5$ BayesMGD and MGD perform similar, while for $\eta = 0.05$ BayesMGD outperforms
MGD. Note that in the given case for $\eta = 0.05$ the number of sample points is smaller than the number 
of parameters, i.e.\ even a linear fit is not fully specified.

\begin{figure}
    \centering
    \includegraphics{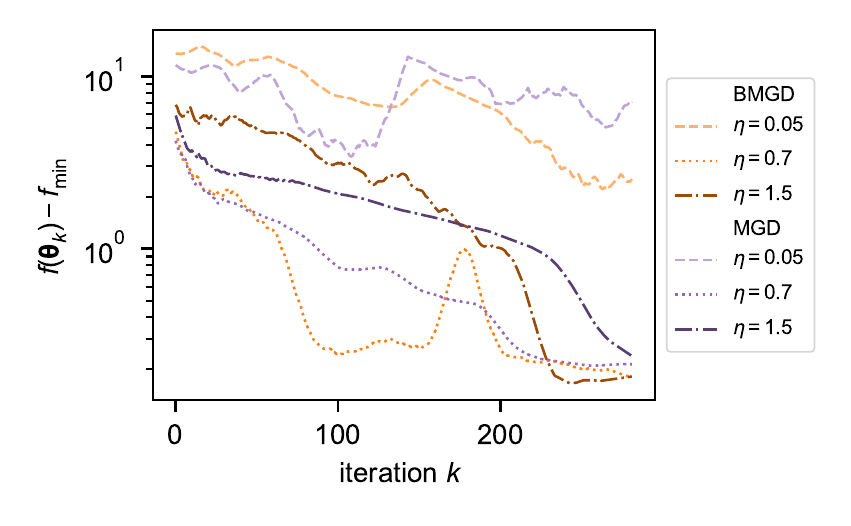}
    \caption{
        Comparison of B(ayes)MGD and MGD with a simulated cost function with 18 parameters.
        As a cost function we chose the cost functions from Ref.~\onlinecite{bosse21} on $2 \times 6$ qubits 
        with 3 layers. The number of shots per evaluation was scaled with $n_{\mathrm{shots}} \sim \eta^{-1}$,
        making the number of shots per optimiser iteration constant. For clarity, we show only the 
        exact function value at the current $\vec\theta$ instead of the noisy evaluations at the sample
        points. These are typically higher and far more spread out.
    }
    \label{fig:optimiser_comparison_ii}
\end{figure}

\end{document}